\documentclass[preprint,12pt,numafflabel]{elsarticle}

\usepackage{amssymb}
\usepackage{lipsum}
\usepackage{amsmath}
\usepackage{multirow}
\usepackage{xcolor}
\usepackage{eqparbox,color}
\usepackage{hyperref}
\usepackage{xcolor,soul,framed} %,caption
\usepackage{xcolor}
\usepackage{fancyhdr}
\journal{ $\begin{array}{l}
     \text{\color{red} This paper appears in the journal ``Computers in Biology and Medicine''.} \\
      \text{\color{red} \text{DOI:~}\href{https://doi.org/10.1016/j.compbiomed.2023.107834}{https://doi.org/10.1016/j.compbiomed.2023.107834}} 
\end{array}$   }

\begin{document}

\begin{frontmatter}



\title{Deep Learning Innovations in Diagnosing Diabetic Retinopathy: The Potential of Transfer Learning and the DiaCNN Model}


\author[inst1]{Mohamed R. Shoaib}

\author[inst3]{Heba M. Emara}
\author[inst1]{Jun Zhao}
\author[inst3,inst5]{Walid El-Shafai}
\author[inst4]{Naglaa F. Soliman}
\author[inst2]{Ahmed S. Mubarak}
\author[inst2]{Osama A. Omer}
\author[inst3,inst4]{Fathi E. Abd El-Samie}
\author[inst2]{Hamada Esmaiel}

\affiliation[inst1]{organization={School of Computer Science and Engineering, Nanyang Technological University},
            city={Singapore},
            postcode={639798}, 
            %state={State One},
            country={Singapore}}
\affiliation[inst2]{organization={Electrical Engineering Department, Aswan Faculty of Engineering, Aswan University},%Department and Organization
            city={Aswan},
            postcode={81542}, 
            %state={State One},
            country={Egypt}}
\affiliation[inst3]{organization={Department of Electronics and Electrical Communications Engineering, Faculty of Electronic Engineering, Menoufia University},
            city={Menouf},
            postcode={32952}, 
            %state={State One},
            country={Egypt}}
\affiliation[inst4]{organization={Department of Information Technology, College of Computer and Information Sciences, Princess Nourah bint Abdulrahman University},
            city={Riyadh},
            postcode={11671}, 
            %state={State One},
            country={Saudi Arabia}}
\affiliation[inst5]{organization={Security Engineering Lab, Computer Science Department, Prince Sultan University},
            city={Riyadh},
            postcode={11586}, 
            %state={State One},
            country={Saudi Arabia}}
\begin{abstract}
Diabetic retinopathy (DR) is a significant cause of vision impairment, emphasizing the critical need for early detection and timely intervention to avert visual deterioration. Diagnosing DR is inherently complex, as it necessitates the meticulous examination of intricate retinal images by experienced specialists. This makes the early diagnosis of DR essential for effective treatment and the prevention of eventual blindness. Traditional diagnostic methods, relying on human interpretation of these medical images, face challenges in terms of accuracy and efficiency. In the present research, we introduce a novel method that offers superior precision in DR diagnosis, compared to these traditional methods, by employing advanced deep learning techniques. Central to this approach is the concept of transfer learning. This entails using pre-existing, well-established models, specifically InceptionResNetv2 and Inceptionv3, to extract features and fine-tune select layers to cater to the unique requirements of this specific diagnostic task. Concurrently, we also present a newly devised model, DiaCNN, which is tailored for the classification of eye diseases. To validate the efficacy of the proposed methodology, we leveraged the Ocular Disease Intelligent Recognition (ODIR) dataset, which comprises eight different eye disease categories. The results were promising. The InceptionResNetv2 model, incorporating transfer learning, registered an impressive 97.5\% accuracy in both the training and testing phases. Its counterpart, the Inceptionv3 model, achieved an even more commendable 99.7\% accuracy during training, and 97.5\% during testing. Remarkably, the DiaCNN model showcased unparalleled precision, achieving 100\% accuracy in training and 98.3\% in testing. These figures represent a significant leap in classification accuracy when juxtaposed with existing state-of-the-art diagnostic methods. Such advancements hold immense promise for the future, emphasizing the potential of our proposed technique to revolutionize the accuracy of DR and other eye disease diagnoses. By facilitating earlier detection and more timely interventions, this approach stands poised to significantly reduce the incidence of blindness associated with DR, thus heralding a new era of improved patient outcomes. Therefore, this work, through its novel approach and stellar results, not only pushes the boundaries of DR diagnostic accuracy but also promises a transformative impact in early detection and intervention, aiming to substantially diminish DR-induced blindness and champion enhanced patient care.

\end{abstract}




\end{frontmatter}


\section{Introduction}
\label{sec:introduction}
Diabetic retinopathy (DR) is a grave ocular condition that impacts individuals suffering from diabetes, making it the primary cause of vision loss among those aged 25 to 74 in developed countries. Approximately 75\% of individuals with diabetes who have had the condition for over 15 years will develop DR. The main cause of DR and its complications is chronic hyperglycemia, which gradually causes microvascular and neurovascular damage in the retina, leading to the development of vision-threatening disorders such as proliferative diabetic retinopathy (PDR) and diabetic macular edema (DME) \cite{int1}.
DR is a disease that worsens over time, starting with mild visual or eye-related symptoms and potentially leading to blindness. Although treatments for DR can slow its progression and reduce vision loss, they rarely fully restore lost eyesight. Consequently, timely detection and treatment of DR are essential to avoid vision loss. To achieve this, people with diabetes should have periodic eye exams to identify and manage DR and other treatable conditions \cite{int2}.
Timely interventions for DR prevent vision loss and maintain the quality of life for people with diabetes. Consequently, healthcare providers should inform individuals with diabetes about the significance of regular eye checkups and their potential for DR. Moreover, healthcare providers should supervise diabetic patients and refer them to a specialist as needed to ensure the prompt and effective treatment of DR. By identifying and treating DR early on; we can decrease the prevalence of blindness and the associated social and economic expenses \cite{int3}.

Diabetes is a persistent medical condition that disturbs the body's metabolism, resulting in elevated blood glucose levels. Regrettably, this ailment can also harm different organs, such as the eyes. DR is a widely recognized diabetes complication that frequently causes vision issues and blindness in working-age adults worldwide. Therefore, a quick diagnosis and treatment of DR are essential for reducing the risk of visual impairment among people with diabetes \cite{int4}.

Efforts to prevent vision impairments in people with diabetes aim to identify DR early and provide timely and appropriate treatment. Screening for DR is a cost-effective approach for early detection, but it necessitates broad population coverage. According to Kjeldsen et al.\cite{int5}, it was estimated that increasing the participation of individuals with type 1 diabetes in screening and treatment by 10\% could lead to annual cost savings of approximately \$16.5 million. This calculation indicates that increasing the percentage of people with type 1 diabetes who undergo DR screening and treatment is not only cost-effective but also financially advantageous. Hence, it is crucial to establish effective DR screening programs that are widely accessible and affordable to minimize the burden of vision loss associated with diabetes. These initiatives should target high-risk populations, including those with type 1 and type 2 diabetes, and promote regular eye exams to detect DR early. Furthermore, raising awareness of DR and its significance among people with diabetes and healthcare professionals could enhance adherence to screening and treatment guidelines, resulting in better clinical outcomes and cost savings.

Diabetic retinopathy represents a frequent complication arising from diabetes and possesses the potential to lead to vision impairment unless it is identified and addressed in a timely manner. The frequency of dilated eye exams has been a subject of debate, with most studies focusing on individuals with moderate-to-severe nonproliferative diabetic retinopathy (NPDR) to determine the need for treatment. However, there is no consensus on the frequency of exams for individuals without retinopathy or with only a few microaneurysms \cite{int6}. Recent research indicates that yearly screening may not be effective for some individuals with type 2 diabetes, and the screening interval should be extended. The current recommendation for yearly dilated eye exams for people with diabetes assumes that the risk of developing proliferative retinopathy or macular edema is high. However, this assumption may not hold for individuals without retinopathy or with only a few microaneurysms. Thus, it is necessary to identify the risk factors linked with retinopathy development and modify the screening interval accordingly \cite{int7}.

Several research studies have aimed to identify the risk factors associated with diabetic retinopathy. Predictors of retinopathy include the severity and duration of diabetes, glycemic control, blood pressure, and lipid levels. In light of these findings, experts suggest that individuals with type 2 diabetes who are at low risk of developing retinopathy may only need to undergo screening every two years \cite{int8}. The frequency of dilated eye exams should be based on each person's specific risk factors for retinopathy development. Individuals without retinopathy or with only a few microaneurysms may not require yearly screening. A personalized approach to screening can help alleviate the burden on patients and healthcare systems while still ensuring timely detection and treatment of diabetic retinopathy \cite{int9}.

So, Diabetic retinopathy (DR) remains a significant global health challenge, with its incidence steadily increasing in parallel with the rise of diabetes mellitus cases. Despite advancements in medical imaging, the accurate diagnosis of DR remains contingent on the expertise of specialized ophthalmologists, who interpret the detailed retinal images. Over the past decade, numerous studies have demonstrated the potential of machine learning techniques in assisting with medical diagnosis. Specifically, for DR, early attempts employed traditional machine learning algorithms with handcrafted features extracted from retinal images.

With the emergence of deep learning, particularly convolutional neural networks (CNNs), a significant transformation took place. CNNs, due to their ability to learn hierarchical features directly from the data, have achieved significant improvements in DR detection accuracy in recent years. For instance, \cite{int7} developed a deep learning algorithm that showed high sensitivity and specificity in detecting referable DR. Similarly, several studies have integrated pre-trained deep learning models, such as VGG, ResNet, and Inception, to further enhance the diagnostic capabilities for DR.

However, even with these advancements, there's room for further innovation, particularly in leveraging the full potential of transfer learning and developing domain-specific architectures tailored for retinal images. This is the gap our research aims to address by introducing the DiaCNN model and harnessing the capabilities of transfer learning with established models like InceptionResNetv2 and Inceptionv3.

The objective of this study is to introduce an accurate diagnostic approach for eye diseases, particularly diabetic retinopathy, using deep learning models. To achieve this, the study utilizes three distinct models, namely Inception-ResNet-v2 and Inception-v3, as transfer learning frameworks for the multi-classification of eye diseases. Moreover, the proposed approach introduces a new model called DiaCNN, which is a residual-based CNN model with skip connections that are specifically designed to diagnose diabetic retinopathy. In the proposed approach, certain layers of the pre-trained models are kept frozen, and their coefficients are utilized as they are, while the last three layers are updated to adapt to the new task of diabetic retinopathy diagnosis. This process ensures that the pre-trained models' learned features are utilized for eye disease diagnosis, and the new model's updated layers help improve the diagnostic accuracy for diabetic retinopathy. The novelty and main contributions of this work are summarized as follows:

\begin{itemize}
\item While various deep learning models have been used for DR detection, our research presents the DiaCNN model, which has been specifically tailored for the classification of eye diseases. This model, through its architecture and design, has demonstrated unparalleled precision in our experiments.
\item Proposing a novel deep learning model called DiaCNN for multi-class classification of eye diseases, particularly diabetic retinopathy diagnosis.
\item Our study uniquely employs transfer learning not just for feature extraction but also for fine-tuning select layers of pre-existing models. By adjusting only the most pertinent layers to the DR diagnostic task, we ensured an optimal balance between leveraging learned features and adapting to the new dataset.
\item Our work does not only rely on a singular model but integrates the strengths of both InceptionResNetv2 and Inceptionv3, coupled with our novel DiaCNN model. This holistic approach ensures robustness and offers multiple avenues for practical deployment.
\item To the best of our understanding, this research is among the limited studies that validate DR diagnostic methods using the diverse Ocular Disease Intelligent Recognition (ODIR) dataset. This not only tests the models against DR but also against other eye diseases, ensuring a comprehensive assessment of their capabilities.
\item Utilizing transfer learning with two pre-trained models, Inception-ResNet-v2 and Inception-v3, for feature extraction and adapting to the new task of diabetic retinopathy diagnosis.
\item Fine-tuning the last three layers of the pre-trained models to enhance their performance for diabetic retinopathy diagnosis.
\item Investigating the performance of DiaCNN and transfer learning-based methods on the Ocular Disease Intelligent Recognition (ODIR) dataset, which includes eight categories: Normal (N), Diabetic Retinopathy (D), Glaucoma (G), Cataract (C), Age-related Macular Degeneration (A), Hypertension (H), Pathological Myopia (M), and miscellaneous diseases/abnormalities (O).
\item Achieving high accuracy in training, testing, and validation for both transfer learning-based models and the proposed DiaCNN model.
\item Comparing the performance of the proposed DiaCNN model with the transfer learning-based models and demonstrating its superiority in terms of accuracy and sensitivity for diabetic retinopathy diagnosis.
\item Presenting an analysis of the performance of the proposed models across various categories of eye diseases, potentially contributing to the early identification and therapeutic intervention for these conditions.
\item Our models have consistently outperformed many state-of-the-art diagnostic methods in terms of accuracy, with DiaCNN achieving a remarkable 98.3\% accuracy in testing. This level of precision represents a significant stride forward in the domain of DR diagnosis.
\end{itemize}

In the realm of deep learning and image classification, the choice of architecture is pivotal. We elected to utilize ResNet-20 primarily due to its architectural simplicity which serves as an advantage in terms of computational efficiency. This is especially beneficial when resources are a limiting factor. Furthermore, a significant feature of the ResNet-20 architecture is the presence of residual connections. These connections have been empirically demonstrated to counteract the vanishing gradient problem, an issue that's often prevalent in deeper networks. Given the intricate nature of retinal images, and the essentiality of preserving minute details for accurate DR diagnosis, ResNet-20, with its ability to learn deep features without gradient diminishment, emerged as a fitting choice for our study.

Our decision to integrate InceptionV3 and InceptionResNetV2 into our methodology was rooted in their proven prowess in prior image recognition tasks. These architectures, inherently designed to capture spatial hierarchies, have the innate capability to discern features across varying scales. Given that retinal images exhibit a multitude of features, ranging from minute vascular anomalies to larger retinal deformities, an architecture that can concurrently focus on different scales is invaluable. Combining the multi-scale feature capturing ability of Inception architectures with the depth of ResNet-20, we believed, would present a comprehensive approach. This integration was intended to leverage the individual strengths of each architecture, ensuring that our model remains sensitive to the intricate features crucial for DR diagnosis.

The methodological crux of our research hinges on the effective fusion of multiple architectures. To elucidate, we employed the Inception modules—InceptionV3 and InceptionResNetV2—as feature extractors. Once these features were extracted, they were channeled into the ResNet-20 architecture for the final classification. The integration process was facilitated by a fusion mechanism. Herein, the extracted features from the Inception modules were concatenated, creating a composite feature vector. This vector was subsequently introduced into the ResNet-20 model. In doing so, we ensured that our model not only captures the depth and intricacy of retinal images but also classifies them with a high degree of accuracy. This fusion, we posit, strikes a balance between depth and breadth, allowing for a nuanced yet comprehensive analysis of retinal images.

The rest of this paper is organized as follows. In Section 2, we delve into the related work, providing a comprehensive overview of existing methods and techniques in the realm of DR diagnosis using deep learning, and emphasizing the gaps that our research aims to address. Section 3 outlines the methodology utilized in this study, including the dataset description, pre-processing steps, and proposed models. Section 4 presents the experimental findings obtained by comparing the performance of the DiaCNN model against the Inception-ResNet-v2 and Inception-v3 models. Section 5 offers a comprehensive analysis and interpretation of the results. Lastly, Section 6 draws conclusions from our research findings, highlighting the implications and potential impact of our work in the field of ophthalmology. We also touch upon potential future directions for further research.

\section{Related Works}
Numerous methods have emerged for the identification of Diabetic Retinopathy (DR) from Ocular Disease Image Repositories (ODIR), encompassing both machine learning and deep learning systems. Image processing techniques play a pivotal role in the detection of DR, and numerous investigations have underscored the efficacy of ODIR datasets in this context. Researchers have introduced diverse methodologies grounded in machine learning, deep learning, and image processing to tackle the challenge of DR detection.

In one study \cite{b1i}, the authors proposed a diagnostic framework for DR using biomarker activation maps (BAM). While deep learning models leveraging optical coherence tomography (OCT) and optical coherence tomography angiography (OCTA) demonstrate impressive diagnostic accuracy for diabetic retinopathy (DR), their decision-making processes often lack interpretability. To tackle this challenge, the BAM (Biomarker-Emphasizing Adversarial Mapping) framework employs generative adversarial learning techniques to create maps that accentuate the key biomarkers employed by the classifier. In a comprehensive evaluation, the authors applied this framework to a dataset comprising 456 macular scans with diagnoses ranging from non-referable to referable DR. The resulting BAMs offer insightful insights into pathological features such as nonperfusion areas and retinal fluid accumulation. These distinctive features hold significant potential for aiding clinicians in enhancing both the precision and comprehensibility of automated DR diagnoses.

In another study \cite{b2i}, the authors introduced an approach to diagnosing DR using digital fundus images. The study aimed to identify critical DR features such as exudates and blood vessels. The authors proposed a method that involved multiple thresholding and morphological operations in segmenting the blood vessels. In a similar fashion, exudate detection involved the application of k-means clustering and contour detection on the original images. The research also delved into strategies like noise reduction to mitigate false positives in vessel segmentation outcomes, along with the utilization of k-means clustering and template matching for optic disc localization. Moreover, the authors introduced a Deep Convolutional Neural Network (DCNN) architecture, comprising 14 convolutional layers and 2 fully connected layers, for automated binary diagnosis of DR. The findings of the study demonstrated promising results, with 95.93\% accuracy in vessel segmentation, 98.77\% accuracy in optic disc localization, and a 75.73\% success rate in DCNN-based diagnosis.

In their study, Gharaibeh and team presented an approach for detecting diabetic retinopathy in retinal fundus images \cite{br1}. This automated diabetic retinopathy screening system involves crucial stages including preprocessing, optic disc detection and elimination, blood vessel segmentation and removal, fovea exclusion, feature extraction, feature selection, and classification \cite{br1}. By implementing this comprehensive workflow, their method facilitates efficient image processing and precise identification of diabetic retinopathy in retinal fundus images. Their approach demonstrated an average accuracy of 98.4\%.

The paper \cite{b3i} introduces a method to diagnose diabetic retinopathy (DR) from fundus images with high accuracy. Their approach involves creating custom convolutional neural network (CNN) models, which can help identify DR early on and prevent blindness. While manual screening can be slow and subjective, and existing machine learning and deep learning methods often rely on pre-trained models or brute-force techniques, the authors' method takes into account the complexity of fundus images.
The researchers proposed a novel method to automatically determine the dimensions of a lightweight CNN tailored for detecting DR lesions in fundus images. This technique integrates k-medoid clustering, principal component analysis (PCA), and evaluations of inter-class and intra-class variabilities. The resulting custom-designed models encode the distinctive features of DR lesions and adapt well to the internal structures of fundus images. The authors evaluated their novel approach using three distinct datasets: one locally sourced from King Saud University Medical City and two established benchmark datasets from Kaggle, specifically, EyePACS and APTOS2019. Their custom-designed models outperformed widely recognized pre-trained CNN models like ResNet152, DenseNet121, and ResNeSt50, all while utilizing fewer parameters. Furthermore, their approach demonstrated performance comparable to the most advanced CNN-based DR screening techniques currently available. In summary, the authors emphasize the substantial potential of their method in DR screening across diverse clinical settings, providing valuable support in identifying patients who may require further assessment and specialized ophthalmological treatment.

In their study \cite{br2}, Nasr et al. addressed the identification of exudates and cotton wool spots in diabetic retinopathy. The disease detection process encompasses various stages, such as initial data processing, optic disc identification and exclusion, vessel segmentation, feature extraction, feature selection, and classification. The main challenge in DR lies in the detection and elimination of the optic disc, which complicates the identification of lesions. To evaluate the effectiveness of their proposed method, the researchers conducted tests using publicly available databases such as DIARETDB0 and DIARETDB1.  The proposed system offered an automated approach for disease detection, which yields promising results in terms of optic disc localization and classification. The system exhibits an impressive sensitivity rate of 99\%, accompanied by a matching specificity rate of 99\%, resulting in an overall accuracy of 98.60\%. 

In their study \cite{b4i}, the authors proposed a technique to enhance medical diagnosis using deep learning interpretability. They aimed to explain the underlying pathological reasons for diabetic retinopathy (DR) predictions by identifying and separating the activation patterns of neurons on which the predictions relied. To achieve this goal, they introduced new pathological descriptors that represent the spatial and appearance characteristics of lesions through the activated neurons of the DR detector. Additionally, they introduced Patho-GAN, a novel network that generates retinal images with medically plausible symptoms by visualizing the symptoms encoded in the pathological descriptor. By manipulating these descriptors, the authors could control the position, quantity, and categories of generated lesions. The generated images had symptoms directly related to DR diagnosis and were superior to previous qualitative and quantitative methods. The authors also highlighted the potential of their method for data augmentation, as it was faster than existing methods and could help in understanding disease progression and training machine learning models. Overall, this approach provides an explanation for DR detector predictions and generates medically plausible retinal images, which could be useful for medical professionals and researchers.

In \cite{br3} Authors present an approach that combined image processing and artificial intelligence to effectively detect diabetic retinopathy in Fundus images while satisfying the desired performance metrics. Our proposed method incorporates a multi-stage automatic detection process. 
The experimental evaluation encompassed various types of diabetic retinopathy, including exudates, micro-aneurysms, and retinal hemorrhages. Through rigorous analysis and comparison, their approach demonstrated detection accuracies surpassing 98.80\%.
In their work \cite{b5i}, researchers introduced an explanatory Artificial Intelligence (XAI) framework known as ExplAIn, designed for the purpose of classifying the severity of Diabetic Retinopathy (DR) using Color Fundus Photography (CFP) images. The system's algorithm effectively segments and categorizes lesions within the images, ultimately leading to the final image-level classification. What sets ExplAIn apart is its ability to provide explanations, a feature not commonly found in black-box AI systems. This XAI architecture is trained end-to-end through image supervision. Additionally, it employs self-supervised techniques to distinguish between foreground and background elements, thereby improving lesion localization and transforming obscured foreground pixels into a more visually comprehensible representation. The authors evaluated ExplAIn on various CFP image datasets at both the image and pixel levels and expect that it will facilitate AI deployment by offering both high classification performance and explainability.

In \cite{b2}, Neelu K. and S. Bhattacharya proposed a technique for the timely detection of diabetic retinopathy (DR) using a deep learning (DL) model based on the PCA-Firefly algorithm. They utilized the Messidor 64k-images dataset from the UCI-ML repository, which includes three categories: No DR, NP-DR, and P-DR. The method relies on a deep neural network (DNN) approach combined with PCA Firefly and Adam Optimizer algorithms. They also incorporated image processing methods, such as image augmentation, rotation, and edge detection. The technique achieved an accuracy of 96\%, a sensitivity of 90\%, and a specificity of 94\%. Although the DNN-PCA-Firefly technique offers better performance, one potential drawback is reduced efficiency due to the application of PCA on DNN and ML.

A project called EviRed was presented in \cite{n1}, which seeks to enhance the screening, diagnosis, and management of DR using AI. DR is a significant cause of blindness in developed nations, and the current classification based on traditional fundus photography offers limited predictive accuracy. The research aims to explore the fusion of different modalities, such as 3-D structural optical coherence tomography (OCT), 3-D OCT angiography (OCTA), and 2-D Line Scanning Ophthalmoscope (LSO), all acquired concurrently using a PLEXElite 9000 device. This fusion is directed towards enhancing the automated detection of proliferative diabetic retinopathy (DR). The ultimate goal is to seamlessly integrate the extensive dataset generated by these modalities with the patient's other medical information to enhance the precision of DR diagnosis and prognosis. This, in turn, empowers ophthalmologists to make more informed decisions in the course of DR follow-up.

Soham et al. \cite{n2} proposed an approach to enhance the robustness of DR feature extraction from digital fundus images. In their study, the authors applied k-means clustering and contour detection techniques to effectively segment blood vessels and exudates, simultaneously mitigating noise. For optic disc localization, a combination of k-means clustering and template matching was employed. The research leveraged a Deep Convolutional Neural Network (DCNN) model consisting of 14 convolutional layers and 2 fully connected layers for the automated binary diagnosis of Diabetic Retinopathy (DR). The proposed approach achieved impressive accuracy rates: 95.93\% for vessel segmentation and 98.77\% for optic disc localization. The DCNN model achieved an accuracy rate of 75.73\%.

Zang et al. \cite{n3} introduce a novel framework called the Biomarker Activation Map (BAM) to facilitate the interpretation of deep learning classifiers in diabetic retinopathy (DR) diagnosis. Using optical coherence tomography (OCT) and its angiography (OCTA), the BAM framework allows clinicians to better understand and verify the decision-making process of deep learning classifiers. The BAM is generated by combining two U-shaped generators that highlight the classifier-utilized biomarkers, producing a different image than the input of the main generator. By generating BAMs, clinicians can improve their utilization and validation of automated DR diagnosis. Based on current clinical standards, the study included 456 macular scans graded as non-referable or referable DR.

Luo and Ye \cite{b3}  propose an automated DR detection strategy based on a Binocular Siamese-like CNN model using the EyePACS 35k images dataset, which includes five classes. They utilize the Inceptionv3 algorithm along with image processing techniques such as Scaling, Normalization, and High-pass processing, achieving an accuracy rate of 94\%, as well as a Sensitivity of 82\% and a Specificity of 70\%. This approach can potentially improve the efficiency of DR diagnosis and screening rates. However, one potential disadvantage is the difficulty of training and testing datasets with paired fundus images.

Meanwhile, et al. \cite{b4} propose an approach for DR detection using an enhanced rider optimization algorithm equipped with deep learning on the DIRECTED B1 images dataset, which contains four classes. Their approach involves employing the DBN methodology in combination with the MGS-ROA algorithm. They also apply various image processing methods, including the conversion of RGB images to the green channel and image enhancement through CLAHE. This approach achieves an accuracy of 93.1\%, with a sensitivity of 86.3\% and a specificity of 95.4\%. Although this strategy employs a superior algorithm for computing accuracy, the disadvantage is that DBNs do not account for the 2D structure of input images, limiting their performance in computer vision applications.

Pao et al. \cite{b5} propose a Bichannel Convolutional Neural Network-based approach to detect DR. The Kaggle dataset images dataset, which includes three classifications (No DR, Mild DR, and Severe DR), is used. The CNN methodology is employed with the Bichannel CNN model, and an image processing technique that resizes and converts the luminance from RGB is used. The technique achieves an accuracy of 87.8\%, with a sensitivity of 77.8\% and a specificity of 93.88\%. The proposed strategy offers several advantages, including improved accuracy and sensitivity, as well as advanced detection of referable DR. The disadvantage is that the input data must be properly pre-processed.
In another study, Cheruku et al. \cite{b6} conducted an experimental study to classify diabetes using a Radial Basis Function Network. The authors employed the Pima Indians Diabetes (PID) dataset and assessed the proposed system's performance using various validity indices: the conventional RBFN, RBFN + Ratio Index, RBFN + Dunn Index, and RBFN + DV Index. The accuracy rates obtained were 68.53\%, 70\%, 69.33\%, and 69.56\%, respectively.

It is noticed from the presented related studies that the previous methods for DR diagnosis using deep learning models have shown promising results, but the performance of these models is often limited by the availability of large annotated datasets and the need for significant computational resources. In addition, transfer learning has emerged as a promising approach for addressing the limitations of deep learning models in medical image analysis. However, few studies have investigated the use of transfer learning in DR diagnosis, and there is a need for more research in this area to improve the accuracy of DR diagnosis. Thus, the limitations of previous works motivated us to present the proposed work in this study. The proposed work aims to address the challenges of DR diagnosis by proposing a transfer learning-based framework that can leverage pre-trained deep learning models to improve the accuracy of DR diagnosis while addressing the limitations of previous methods.

\section{Materials and Methods}
\subsection{Dataset Description}
\begin{figure}[htbp]
%\label{fig: loss of best model}
\centerline{\includegraphics[width=.7\textwidth]{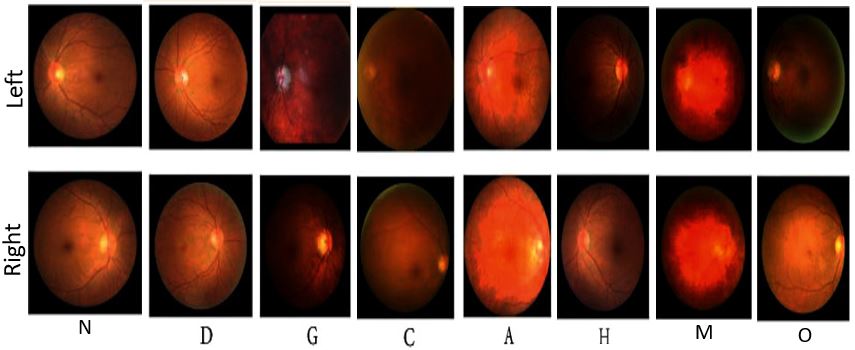}}
\caption{Samples of ODIR dataset used for DR detection.}
\label{fig1}
\end{figure}

\begin{figure}[htbp]
%\label{fig: loss of best model}
\centerline{\includegraphics[width=.7\textwidth]{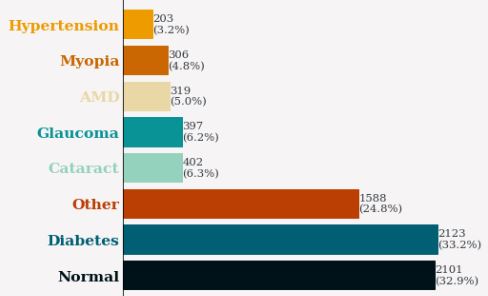}}
\caption{Visualization of the dataset in the form of a bar chart.}
\label{bar1}
\end{figure}

In this study, the ODIR (Ocular Disease Intelligent Recognition) dataset \cite{b1} was utilized. This dataset, available on Kaggle, is widely regarded as a comprehensive resource for the detection of eye diseases. It consists of fundus images that are categorized into eight classes based on ocular disease classification. The dataset consists of 5000 color fundus photographs, categorized into various classes, namely normal (N), myopia (M), hypertension (H), diabetes (D), cataract (C), glaucoma (G), age-related macular degeneration (A), and other abnormalities or diseases (O). These images have been divided into separate training and testing subsets for analysis. The training subset encompasses a little over 4000 cases, while the remaining cases are allocated to the testing subset. In this study, all images were resized to dimensions of $224\times224$ pixels.
For more extensive information on the image distribution within the ODIR dataset, please consult the details provided in Table \ref{tdata}. Moreover, you can view sample images from the dataset in Figure \ref{fig1}. Further insight into the image distribution can be found in Figure \ref{bar1}, which presents a bar chart. The horizontal axis depicts the patient count, while the vertical axis represents the various disease categories. The chart illustrates the distribution of training cases across each class within the dataset. Based on the chart, the normal (N) class shows the highest patient case count (1135), with the diabetes (D) class following closely. In contrast, the hypertension (H) class has the fewest patient cases.
 Moreover, as illustrated in Figure \ref{fig1}, you can observe sample fundus images from the dataset. The terms "left" and "right" specify whether the image corresponds to the left or right eye, respectively.


\begin{table}[]
\centering
\caption{Image distribution within the ODIR dataset.}
\label{tdata}
\begin{tabular}{|c|c|c|}
\hline
\textbf{No. of classes} & \textbf{Training cases} & \textbf{Labels} \\ \hline
\textbf{1}   & \textbf{1135}     & \textbf{Normal (N)}      \\ \hline
\textbf{2}   & \textbf{1131}     & \textbf{Diabetes (D)}      \\ \hline
\textbf{3}   & \textbf{207}      & \textbf{Glaucoma (G)}      \\ \hline
\textbf{4}   & \textbf{211}      & \textbf{Cataract (C)}      \\ \hline
\textbf{5}   & \textbf{171}      & \multicolumn{1}{l|}{\textbf{Age-related macular (A)}}      \\ \hline
\textbf{6}   & \textbf{94}       & \textbf{Hypertension (H)}      \\ \hline
\textbf{7}   & \textbf{177}      & \textbf{Myopia (M)}      \\ \hline
\textbf{8}   & \textbf{944}      & \textbf{Other diseases (O)}      \\ \hline
\end{tabular}
\end{table}

\subsection{Proposed Method for DR Detection }
In this section, we present DiaCNN, a deep-learning model based on the ResNet-20 architecture for image classification. Although ResNet-20 has been successful in various image classification tasks, we aimed to enhance its performance and investigate transfer learning-based models. Therefore, we incorporated InceptionV3 and InceptionResNetV2 and compared their performance with DiaCNN. Transfer learning is a valuable approach that enhances model performance on limited datasets by leveraging pre-trained models trained on extensive datasets. This investigation aims to assess the efficacy of DiaCNN while also conducting a performance comparison with contemporary transfer learning-based models.

\begin{figure}[htbp]
%\label{fig: loss of best model}
\centerline{\includegraphics[width=.7\textwidth]{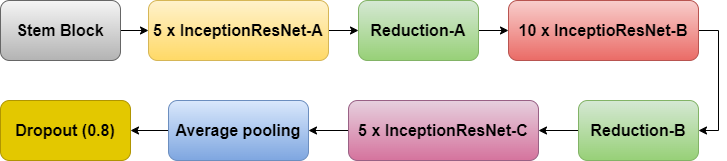}}
\caption{The basic architecture of InceptionResnetv2.}
\label{fig2}
\end{figure}

Deep learning has found a wide range of applications, such as speech recognition, medical data categorization, and lesion detection through segmentation. This research proposes the use of InceptionResNetV2 and InceptionV3 as pre-trained CNN models to identify DR \cite{b7,b9}. The block diagram in Figure \ref{fig2} illustrates this method. In a standard CNN architecture, various components are employed, comprising an input layer, convolutional layers, pooling layers, fully connected layers, and an output layer \cite{b8, b9, b10, b11, b12, b13, b14, b15, b16, b17, b18, b19, b20, b21, b22,b23, b24, b25, b26,bh,bh1, b27, b28, b29, b30}. The construction of a pre-trained CNN model follows these principles:

\begin{itemize}
\item Input layer: The model takes an input image scan with a resolution of $224\times224$ pixels.
\item Convolutional layers: Incorporates a convolutional layer (Conv), followed by a Batch Normalization (BN) layer and a Rectified Linear Unit (ReLU) layer. The convolutional layer compresses image features by applying three convolutions to input images, each using distinct filters with a consistent window size of 3. The filter sizes are 8, 16, 32, 64, and 128 for the first through fifth convolutions. To improve test accuracy and prevent overfitting, we incorporate Batch Normalization layers during optimization. In the training phase, ReLU activation functions are employed to introduce element-wise non-linearity in the model.
\item Pooling layer: This is an essential component for feature extraction in deep learning. It is responsible for identifying significant features of each map. To implement this layer, the max-pooling method is used, which generates a feature vector of fixed length by combining the max-pooled vectors. In our implementation, a stride of 2 is set with a max-pooling window size of $2\times2$.
\item Fully connected layers: The provided function accepts a basic vector as its input and generates a solitary output vector. To achieve this, we utilize a paradigm consisting of four fully connected (FC) layers. The output layer is fully connected and employs SoftMax activation, enabling the classification of input images into one of four categories.
\end{itemize}

\begin{figure}[htbp]
\label{fig: loss of best model}
\centerline{\includegraphics[width=.7\textwidth]{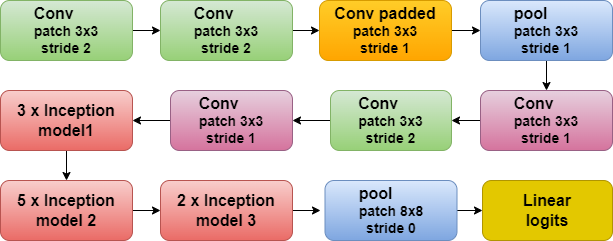}}
\caption{The basic architecture of Inception-v3.}
\label{fig3}
\end{figure}

InceptionResNetv2 is a type of convolutional neural network that enhances the Inception family by incorporating residual connections. In this network, residual functions learn from layer inputs through skip connections and replace the filter concatenation step of the Inception architecture. The InceptionResnetv2 \cite{b13} combines the Inception-Resnet-v2 structure with Residual Connections. The Inception-Resnet block \cite{b13} merges multiple-sized convolutional filters via residual connections. By using residual connections, InceptionResNetv2 addresses the degradation issue and reduces training time caused by deep structures. The basic network topology of InceptionResNetv2 is illustrated in Figure \ref{fig3}.

The mathematical analysis of InceptionResNetv2 involves examining the architecture and parameters of the model \cite{b10, b13, b29, b30, b31, b32, b33, b34, bb35, bb36}. Here are some key equations that govern the model's behavior:

Convolutional layer equation: The convolutional layer's output is generated by convolving the input image $(x)$ with a set of filters $(w)$, incorporating a bias term $(b)$, and subsequently applying an activation function $(\sigma)$:
\begin{equation}
y = \sigma(\text{conv}(x, w) + b)
\end{equation}
Here, $\text{conv}$ is the convolution operation, and the + operator denotes element-wise addition.

Pooling layer equation: The output of a pooling layer is the maximum (or average) value of a window of pixels in the input image:
\begin{equation}
y = \text{maxpool}(x)
\end{equation}
Here, $\text{maxpool}$ is the maximum pooling operation.

Inception module equation: The output of an Inception module is the concatenation of the outputs of several convolutional and pooling layers, followed by a 1x1 convolutional layer to reduce the number of channels:
\begin{equation}
\scalebox{0.85}{$y$ = $\text{conv1}$($\text{concat}$($\text{conv1}$($x$), $\text{conv3}$($x$), $\text{conv5}$($x$), $\text{maxpool}$($x$)))}
\end{equation}
Here, $\text{conv1}$, $\text{conv3}$, and $\text{conv5}$ are convolutional layers with filter sizes of $1\times1$, $3\times3$, and $5\times5$, respectively, and $\text{maxpool}$ is a pooling layer. The $\text{concat}$ function combines the outputs of these layers along the channel dimension.

Residual connection equation: The output of a residual connection is the sum of the input features $(x)$ and the output of another layer $(y)$:
\begin{equation}
y = x + F(x)
\end{equation}
Here, $F(x)$ is the output of a convolutional layer that operates on $x$.

Stem layer equation: The stem layer of InceptionResNetv2 performs initial feature extraction by applying a series of convolutional and pooling layers:
\begin{equation}
y = \text{maxpool}(\text{conv}(x, w1)) + \text{maxpool}(\text{conv}(x, w2))
\end{equation}
Here, $w1$ and $w2$ are sets of filters with different sizes.

Reduction module equation: The reduction module's output is obtained by combining the results from multiple convolutional and pooling layers. This is followed by the application of a $1\times1$ convolutional layer, which serves to decrease the channel count:
\begin{equation}
\scalebox{0.85}{$y$ = $\text{conv1}$($\text{concat}$($\text{maxpool}$($x$), $\text{conv1}$($x$), $\text{conv3}$($x$), $\text{conv5}$($x$)))}
\end{equation}
Here, $\text{maxpool}$, $\text{conv1}$, $\text{conv3}$, and $\text{conv5}$ are operations as defined previously.

These equations describe the basic operations of the InceptionResNetv2 model. Throughout the training process, the model fine-tunes its parameters, such as weights and biases, to minimize the disparity between its predicted outputs and the real labels associated with the training dataset. This process is typically carried out using backpropagation and stochastic gradient descent algorithms.

Inceptionv3 \cite{b14} is a CNN architecture that builds upon the successful GoogLeNet \cite{b15} model, which has shown high accuracy in classifying biomedical data using transfer learning \cite{b16, b17, bh2}. In addition, inception-v3 introduces an inception module that combines convolutional filters of various sizes into a single filter, similar to GoogLeNet. This reduces the number of learnable parameters, resulting in lower computational complexity.

%\begin{figure}[htbp]
%\label{fig: loss of best model}
%\centerline{\includegraphics[width=1\columnwidth]%{figures/Picture5.jpg}}
%\caption{The basic architecture of Inception-v3.}
%\label{figben}
%\end{figure}

The mathematical analysis of the Inception-v3 architecture involves studying its parameters and operations \cite{b10, b13, b29, b30, b31, b32, b33, b34}. Here are some of the key equations that govern its behavior:

\begin{equation}
y = \sigma(\text{conv}(x, w) + b)
\end{equation}

The output of a convolutional layer can be obtained by convolving the input image $(x)$ with a set of filters $(w)$ and adding a bias term $(b)$. The $\sigma$ function represents the activation function, such as ReLU, and the $+$ operator denotes element-wise addition.

\begin{equation}
y = \text{maxpool}(x)
\end{equation}

The output of a pooling layer can be obtained by taking the maximum (or average) value of a window of pixels in the input image, where $\text{maxpool}$ is the maximum pooling operation.
\begin{equation}
\scalebox{0.85}{$y$ = $\text{conv1}$($\text{concat}$($\text{conv1}$($x$), $\text{conv3}$($x$), $\text{conv5}$($x$), $\text{maxpool}$($x$)))}
\end{equation}
The Inception module plays a vital role in the Inceptionv3 architecture, enabling the model to capture features across various scales. This module's output is generated by concatenating the results of multiple convolutional and pooling layers. Subsequently, a $1 \times 1$ convolutional layer is applied to reduce the channel count. The $\text{concat}$ function is used to merge the outputs of these layers along the channel dimension. Additionally, we have $\text{conv1}$, $\text{conv3}$, and $\text{conv5}$, which denote convolutional layers with filter sizes of $1 \times 1$, $3 \times 3$, and $5 \times 5$, respectively. Finally, there's $\text{maxpool}$ representing a pooling layer.
\begin{equation}
y = fc(\text{avgpool}(\text{conv}(x)))
\end{equation}
The auxiliary classifier is a secondary classifier added to the network to encourage the model to learn discriminative features. The auxiliary classifier's output can be acquired through a sequence of operations, starting with a convolutional layer, followed by global average pooling (denoted as $\text{avgpool}$), and concluding with a fully connected layer represented as $fc$.

\begin{equation}
y = \text{maxpool}(\text{conv}(x, w1)) + \text{maxpool}(\text{conv}(x, w2))
\end{equation}
The stem layer of Inception-v3 performs initial feature extraction by applying a series of convolutional and pooling layers, where $\text{conv}$ represents the convolution operation, $\text{maxpool}$ represents the maximum pooling operation, and $w1$ and $w2$ represent sets of filters with different sizes.

\begin{equation}
y = \text{conv1}(\text{concat}(\text{maxpool}(x), \text{conv1}(x), \text{conv3}(x), \text{conv5}(x)))
\end{equation}
The output of a reduction module is computed by concatenating the results from multiple convolutional and pooling layers. This concatenation is then followed by a $1 \times 1$ convolutional layer, which serves to reduce the number of channels. It's worth noting that $\text{maxpool}$, $\text{conv1}$, $\text{conv3}$, and $\text{conv5}$ have been defined previously.These equations represent the fundamental operations of the Inceptionv3 architecture. In the training phase, the model fine-tunes its parameters, including weights and biases, with the aim of minimizing the dissimilarity between its predicted results and the real labels of the training dataset. This procedure generally involves utilizing backpropagation and stochastic gradient descent (SGD) techniques.

To provide further clarity and enhance the reproducibility of the study, an elaboration on the fine-tuning process was employed for the InceptionV3 and InceptionResNetV2 models.
In the feature extraction step, the pre-trained InceptionV3 and InceptionResNetV2 models were utilized. These models have been trained on large-scale image datasets, enabling them to learn general visual representations. By removing the last fully connected layers, we retained the convolutional base of these models, which is responsible for extracting meaningful features from images.
Next, in the fine-tuning step,  new fully connected layers are added on top of the convolutional base to adapt the models to our specific ocular disease classification task. The weights of the pre-trained layers were kept frozen during this process to preserve the learned representations. By doing so, we aimed to leverage the knowledge acquired by the models on general image features while allowing the new layers to specialize in capturing disease-specific patterns.
During training, we employed specific model parameters to monitor and save the best-performing model based on validation accuracy. We set the parameters as follows: monitor="val\_accuracy",
save\_best\_only=True, mode="auto", verbose=1. This ensured that we captured the model with the highest accuracy on the validation set for further evaluation and testing.

In terms of hardware, the experiments were conducted on an MSI-ML laptop equipped with an Intel(R) Core(TM) i7-10750H CPU @ 2.60GHz processor, 32.0 GB of installed RAM (31.8 GB usable), and a 64-bit operating system. The laptop also featured an NVIDIA GeForce RTX 2070 graphics card, which accelerated deep learning computations.
For enhancing the training process, we employed the Adam optimizer with its default settings. The Adam optimizer combines the benefits of adaptive learning rates and momentum to efficiently update the model's parameters. Specifically, it utilizes momentum values of 0.9 and 0.999 to control the momentum accumulation and adaptive learning rate scaling.
To further enhance the training process, we utilized additional techniques such as early stopping. This approach serves to mitigate overfitting by continuously assessing the validation loss and halting the training process if there is no discernible improvement for a specified number of epochs. Our early stopping criteria are configured as follows: a factor of 0.3, a patience of 2, a minimum delta of 0.001, mode set to 'auto,' and verbosity level 1.

In this study, we analyze pre-trained models and their image input sizes and training options, which are summarized in Table \ref{t1}. Our training techniques effectively address the problem of deterioration and achieve convergence in a minimum number of iterations. To achieve this, we utilize stochastic gradient descent (SGD) due to its rapid convergence and short running time. Additionally, we apply ReLU activation to all convolutional layers.
\begin{table*}[h]
\centering
\caption{Training options for different pre-trained models.}
\resizebox{\textwidth}{!}{%
\renewcommand{\arraystretch}{2}
\begin{tabular}{|c|c|c|}
\hline
\multicolumn{3}{|c|}{\textbf{Training options (random initialization weights, batch size =64,
learning rate = 0.00001 and number of epochs = 20)}} \\ \hline
\textbf{Model} & Input size & No. of layers \\ \hline
\textbf{InceptionResNetv2 \cite{b13}} & 229 × 229 & 164 \\ \hline
\textbf{Inceptionv3 \cite{b14}} & 229 × 229 & 48 \\ \hline
\end{tabular}
\label{t1}%
}
\end{table*}
\subsection{DiaCNN Proposed Model}
\begin{figure*}[htbp]
\label{fig: loss of best model}
\centerline{\includegraphics[width=0.8\textwidth]{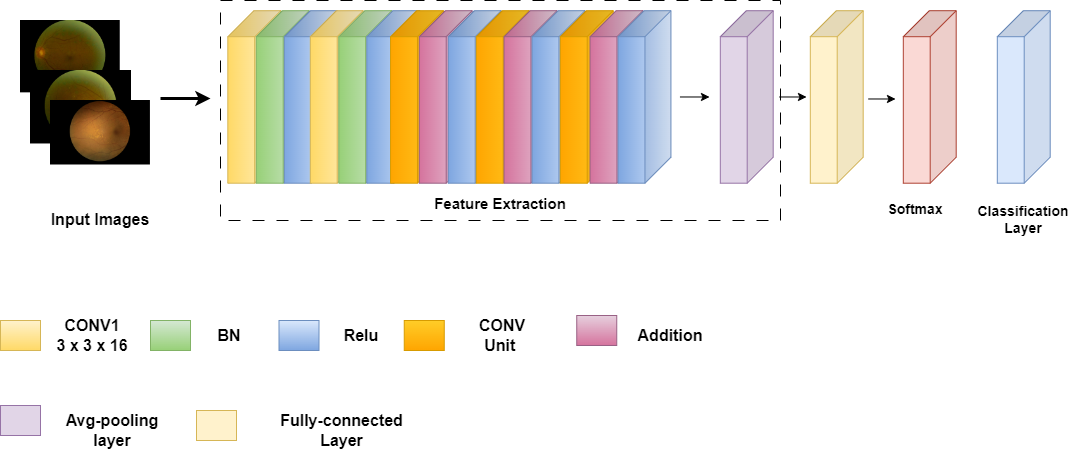}}
\caption{Block diagram of the proposed method for DR detection.}
\label{figben}
\end{figure*}

The suggested design in Figure \ref{figben} incorporates an input layer tailored for a $32\times32\times3$ image, succeeded by a $3\times3$ convolutional layer featuring 16 filters. Padding in the convolutional layer is employed to preserve the original spatial dimensions of the input. Afterward, the output undergoes normalization, achieving zero mean and unit variance through batch normalization. The output of the batch normalization layer undergoes an element-wise ReLU activation function.

The model is composed of multiple residual blocks, each comprising two convolutional layers and an extra layer for element-wise addition of the second convolutional layer's output to the input. The number of filters in each convolutional layer is regulated by the 'netWidth' hyperparameter. Batch normalization is then applied to the output of the second convolutional layer, followed by a ReLU activation layer. Subsequently, a $3\times3$ convolutional layer, with the same number of filters as the initial convolutional layer, is appended. The output of this second convolutional layer is also subjected to batch normalization and added to the input of the residual block. Finally, another ReLU activation layer is applied.

The residual block outputs are fed into a global average pooling layer, which computes the mean value for each feature map across the entire spatial area, producing a $(1\times1\times n)$ tensor. Subsequently, a fully connected layer with two output units, representing the dataset's two classes, is utilized. Finally, the output of the fully connected layer is passed through a softmax activation layer to yield a probability distribution over the classes, achieved by applying the softmax function to the output.

At the end of the network, the softmax activation layer generates class probabilities. A classification layer calculates the cross-entropy loss between predicted and actual class labels, and the weights are updated by backpropagating the error through the network. To improve gradient flow, skip connections that bypass some residual blocks in the network are created using $1\times1$ convolutional layers and batch normalization layers.

In our study, evaluating the precision and robustness of the models is paramount. To ensure a comprehensive evaluation, we employ several performance metrics that have been widely adopted in the domain of medical image analysis. \textbf{\textit{Accuracy:}} Represents the proportion of correctly identified cases out of the total cases. It's a measure of the model's overall effectiveness. \textbf{\textit{Precision:}} Also known as the positive predictive value, it's the proportion of positive identifications that were actually correct. \textbf{\textit{Recall (Sensitivity):}} Represents the ability of the model to identify all relevant instances. \textbf{\textit{F1 Score:}} The harmonic mean of precision and recall. It provides a balance between precision and recall. \textbf{\textit{Area Under the Receiver Operating Characteristic Curve (AUC-ROC):}} Evaluates the performance of the binary classification system and its discriminative power between positive and negative classes. 

The Receiver Operating Characteristic (ROC) curve is a crucial tool for assessing diagnostic tests, commonly utilized in tandem with deep learning models for medical image classification. In a binary classification system, the ROC curve graphically illustrates the model's performance across various classification thresholds. It showcases the True Positive Rate (TPR) or Sensitivity against the False Positive Rate (FPR) or 1-Specificity. The diagonal line in the ROC space, representing an area of 0.5, indicates a classifier that performs no better than random guessing. In contrast, a curve that closely follows the top-left border of the ROC space reflects an excellent classifier, with a true positive rate much higher than the false positive rate across most threshold values.

A valuable metric derived from the ROC curve is the Area Under the Curve (AUC), which quantifies the overall ability of the model to discriminate between positive (in our case, instances of Diabetic Retinopathy) and negative classes. The AUC value ranges between 0 and 1, with a higher value indicating superior model performance. Specifically:
\begin{itemize}
\item An AUC of 1 denotes a perfect classifier, correctly distinguishing between the classes for all threshold values.
\item An AUC close to 0.5 suggests the classifier performs no better than random guessing.
\item An AUC value between 0.5 and 1 indicates varying degrees of classifier performance, with values closer to 1 reflecting higher discriminatory power.
\end{itemize}
For our research, we calculated the AUC values for each of the employed models: InceptionResNetv2, Inceptionv3, and DiaCNN, using the Ocular Disease Intelligent Recognition (ODIR) dataset. The results further validate our model's diagnostic capabilities, offering an additional layer of evaluation beyond mere accuracy metrics. The AUC values, when considered in conjunction with other performance metrics, provide a holistic perspective on the models' robustness and reliability in diagnosing Diabetic Retinopathy.

By employing these metrics, we aim to provide a multifaceted view of our models' capabilities, ensuring that our results are not only high-performing but also consistent across various evaluation parameters.

The deep learning classifier's performance is assessed through various metrics, including sensitivity (Sen), specificity (Spec), accuracy (Acc), precision (Preci), and F1 score \cite{b18}, which are computed based on the confusion matrix. Table \ref{t2} presents the expected outputs in the four quadrants of the confusion matrix. True positives ($T_p$) correspond to the accurately identified anomalous instances, while true negatives ($T_n$) indicate the correctly classified normal instances. False positives ($F_p$) represent normal instances mislabeled as anomalies, and false negatives ($F_n$) represent anomalies misclassified as normal.

\begin{table}[h]
\centering
\caption{Confusion matrix.}
%\resizebox{\textwidth}{!}{%
\renewcommand{\arraystretch}{2}
\begin{tabular}{|c|c|c|}
\hline
 & \textbf{Actually positive (1)} & \textbf{Actually negative (0)} \\ \hline
\textbf{Predicted positive} & $T_{p}$s &$F_{p}$s \\ \hline
\textbf{Predicted negative} & $F_{n}$s & $T_{n}$s \\ \hline
\end{tabular}
\label{t2}
%}
\end{table}

Sensitivity is given by:
\begin{equation}
		Sen=\frac{T_p}{T_p+F_n} \times 100
	\end{equation}
	
	Specificity is given by:
\begin{equation}
		Spec=\frac{T_n}{T_n+F_p} \times 100
	\end{equation}
	
	Accuracy is given by:
\begin{equation}
		Acc=\frac{T_p+T_n}{T_p+T_n+F_p+F_n} \times 100
	\end{equation}

 	F1 score is given by:
\begin{equation}
		F1 score=\frac{T_{p}}{T_{p}+\frac{1}{2}(F_{p}+F_{n})} \times 100
	\end{equation}

The F1 score, or $F$-measure, is a valuable metric for evaluating test accuracy. It's calculated by dividing true positives by all positive results (true and false positives). Recall measures correctly identified positives against total positives. The F1 score is derived from the harmonic mean of precision and recall, as outlined in \cite{b19}.
 
 \section{Results}
 
\subsection{Experimental Setup}
 
\textbf{Dataset Configuration:} We utilized the ODIR dataset. This dataset encompasses retinal images categorized into eight distinct eye disease classifications. The dataset was divided into a 80-10-10 split for training, validation, and testing, respectively.

\textbf{Preprocessing:} All images were resized to a uniform size of 299x299 pixels, as this dimension aligns with the input requirements of the InceptionResNetv2 and Inceptionv3 architectures. Furthermore, image augmentation techniques, such as random rotations, zooming, and horizontal flipping, were employed to increase the diversity of training data and prevent overfitting.

Given the diverse nature and quality of retinal images in the Ocular Disease Intelligent Recognition (ODIR) dataset, it was imperative to apply several preprocessing steps to ensure the consistency and reliability of our model's performance.
\begin{itemize}
\item Image Resizing: To maintain uniform input dimensions for our CNN models, all retinal images were resized to a standard resolution of 224x224 pixels without altering their original aspect ratio.

\item Histogram Equalization: To enhance the contrast of the retinal images and emphasize subtle features, we utilized histogram equalization. This process helps in improving the visibility of blood vessels and microaneurysms, which are vital indicators of Diabetic Retinopathy.

\item Noise Reduction: Given the potential presence of artifacts and noise in some images, we employed Gaussian blurring. This step aids in minimizing the noise while preserving essential details in the image.

\item Augmentation: To bolster the robustness of our model, especially given the limited dataset size, we augmented our dataset by generating variations of the original images. Techniques included random rotations, zooms, and horizontal flips. This not only expanded the effective size of our dataset but also equipped our models to recognize DR indicators under a variety of image conditions.

\item Normalization: Lastly, we normalized the pixel values of each image to the range [0,1], ensuring that the models' training process remains stable and converges faster. This step is pivotal in harmonizing the intensity levels across the dataset.
\end{itemize}

\textbf{Hardware and Software:} The experiments were conducted on a workstation equipped with an NVIDIA RTX 3090 GPU, 64GB RAM, and an Intel Core i9 processor. We employed TensorFlow 2.14 as our deep learning framework.

\textbf{Transfer Learning Models (InceptionResNetv2 and Inceptionv3):} The pre-trained weights from ImageNet were loaded, and the final fully connected layers were fine-tuned for our 8-category classification task. The learning rate was initialized at 1e-4 and decreased by half every five epochs using a learning rate scheduler. The models were trained for a total of 30 epochs.

\textbf{DiaCNN Model:} The model was trained from scratch using the same dataset split. We adopted the Adam optimizer with a learning rate of 1e-3. Given the depth of this model, a longer training duration of 50 epochs was deemed necessary.

Models were evaluated based on their accuracy, precision, recall, F1-score, and ROC-AUC on the testing dataset. We also leveraged the validation dataset to tune hyperparameters and avoid overfitting.

For a comprehensive analysis, our models were juxtaposed against existing state-of-the-art methods on the ODIR dataset, ensuring that all methods were evaluated under equivalent conditions.

 \subsection{Results for Transfer Learning-based Models}
A pre-processing step is applied to the data before feeding it to the CNN model for classification based on the correct class. The performance of the CNN models varies according to their parameters, which are detailed in Tables \ref{t1} and \ref{t3}. These tables list information about each pre-trained model after being utilized as a transfer learning model, including input size, number of layers, total parameters, trainable parameters, and non-trainable parameters. Additionally, the hyperparameters used to train these models, such as learning rate, batch size, and epoch count, are also provided.
\begin{table}[h]
\centering
\caption{The Total number, Trainable and Non-trainable parameters of Transfer learning models.}
\resizebox{\columnwidth}{!}{%
%\resizebox{\textwidth}{!}{%
\renewcommand{\arraystretch}{2}
\begin{tabular}{|c|c|c|c|}
\hline
\textbf{Model} & \textbf{Number of Trainable parameters} & \textbf{Number of Non-trainable parameters} & \textbf{Number of Total parameters} \\ \hline
Transfer learning based on InceptionResNetv2 & 54,722,273 & 60,544 & 54,782,817 \\ \hline
Transfer learning based on Inceptionv3 & 18,096,770 & 229,056 & 18,325,826 \\ \hline
\end{tabular}%
\label{t3}
}
\end{table}

\begin{figure*}[htbp]
%\label{fig: loss of best model}
\centerline{\includegraphics[width=.7\textwidth]{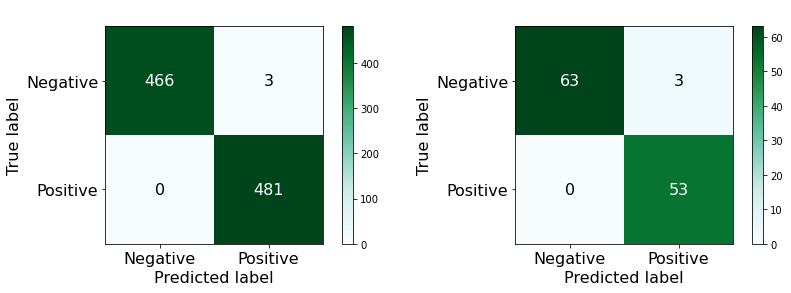}}
\caption{Training (left) and testing (right) confusion matrices of the transfer learning based on InceptionResNetv2.}
\label{con4}
\end{figure*}

\begin{figure}[htbp]
%\label{fig: loss of best model}
\centerline{\includegraphics[width=.5\textwidth]{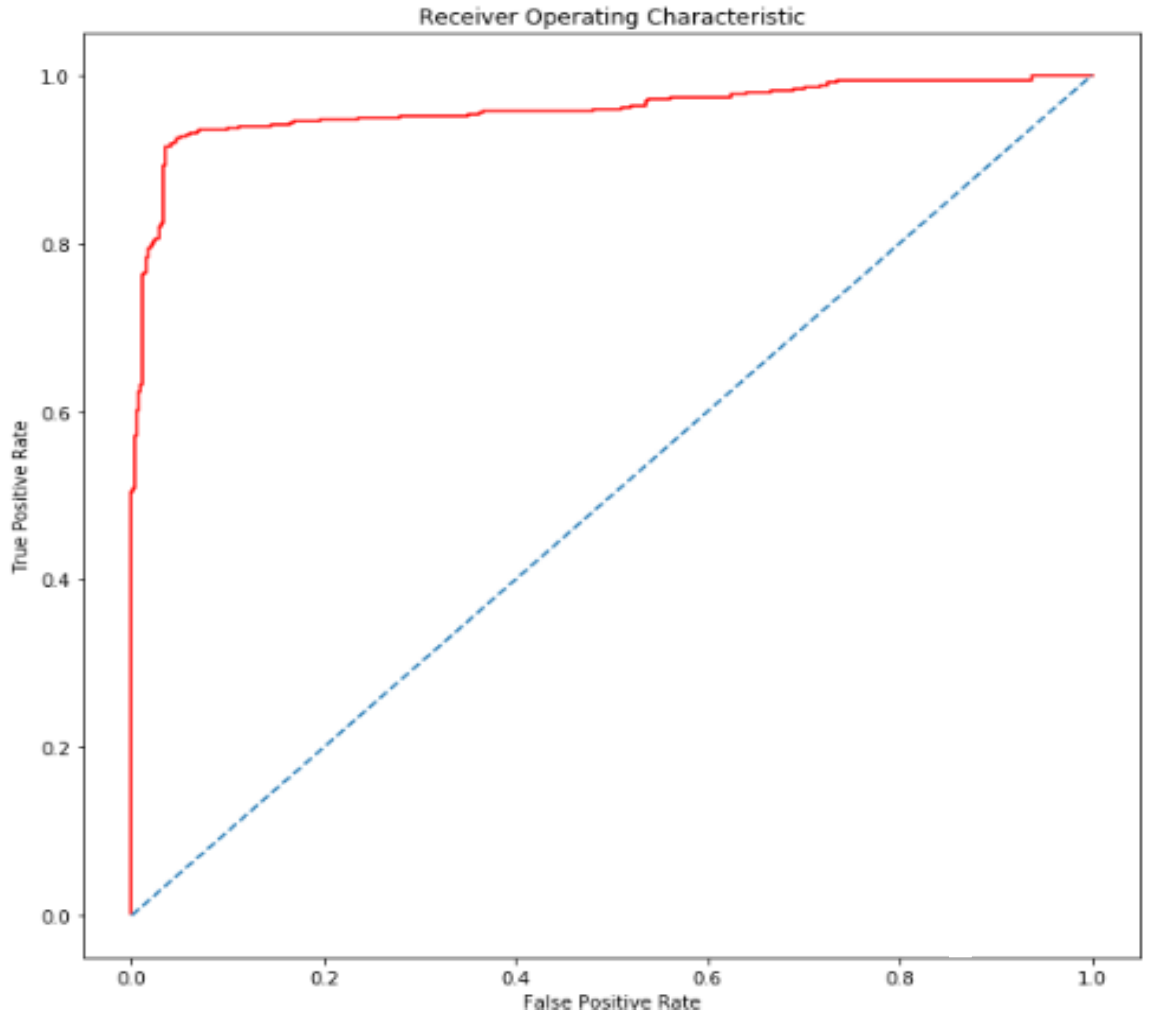}}
\caption{ROC curve for the transfer learning based on InceptionResNetv2.}
\label{r96}
\end{figure}

\begin{figure}[htbp]
%\label{fig: loss of best model}
\centerline{\includegraphics[width=.7\textwidth]{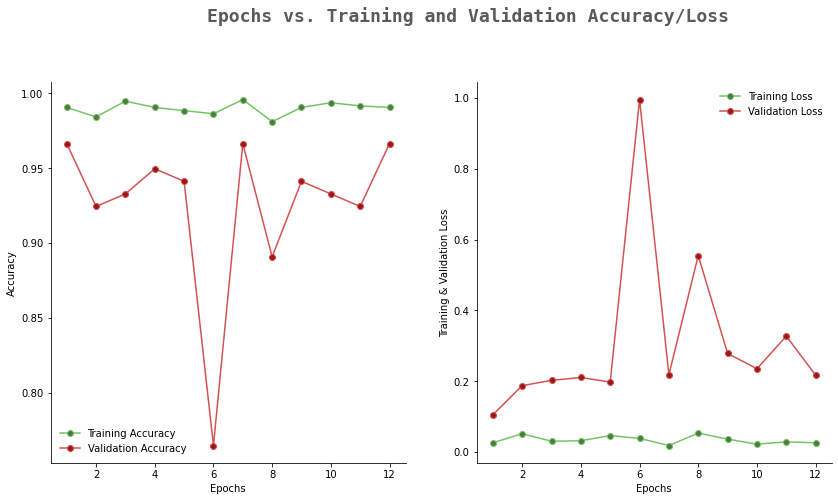}}
\caption{Training progress curve for the transfer learning-based InceptionResNetv2 model.}
\label{tr4}
\end{figure}

\begin{figure}[htbp]
%\label{fig: loss of best model}
\centerline{\includegraphics[width=.7\textwidth]{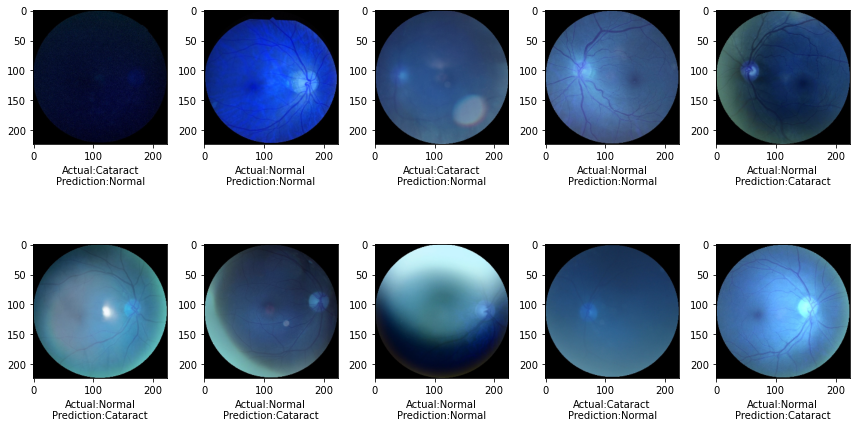}}
\caption{Samples of output results for the transfer learning-based InceptionResNetv2 model.}
\label{res1}
\end{figure}

This paper evaluates the performance of two pre-trained CNN models, namely InceptionResNetv2 and InceptionV3, used as transfer learning-based models. The results of the InceptionResNetv2 model are presented in Table \ref{ta4}, which includes precision, recall, F1 score, and support metrics for each class label. The model achieved an accuracy of 96\%. In terms of evaluating its performance, precision gauges its capacity to accurately identify positive samples, whereas recall measures its capacity to capture all positive samples. The F1 score gives a comprehensive assessment of the model's performance, calculated as the harmonic mean of precision and recall. The support value indicates the sample count in each class. Table \ref{ta4} demonstrates the model's high precision, recall, and F1 score for class labels 0 and 1. Both the macro average and weighted average values for precision, recall, and F1 score were 0.96, indicating the model's robust overall performance. These results suggest that the InceptionResNetv2 model performed well in classifying different retina diseases.
The validation and testing phases yielded confusion matrices, as depicted in Figure \ref{con4}. Additionally, Figure \ref{r96} displays the ROC curve for the transfer learning of the InceptionResNetv2 model. Notably, this model exhibited remarkable performance in correctly identifying diabetic retinopathy cases, characterized by high true positive rates and low false negative rates in both phases. However, the model had a relatively high false positive rate for the validation phase, incorrectly classifying some negative cases as positive. This highlights the need for further refinement and optimization of the model to reduce its false positive rate.
Figure \ref{tr4} presents the training validation progress curve. The model rapidly acquired knowledge from the training dataset, leading to impressive accuracy. However, there was some overfitting observed as the validation accuracy stagnated or even decreased while the training accuracy continued to increase. This indicates that the model may have learned to fit the training data too closely and was not able to generalize well to new data. The loss curve also shows similar behavior, with the training loss decreasing steadily while the validation loss stagnates or even increases. Therefore, it is important to address over-fittings, such as regularization or data augmentation techniques, to improve the model performance on new data. Figure \ref{res1} presents some of the output results.
\begin{table}[]
\centering
\caption{Classification report of transfer learning-based InceptionResNetv2 model.}
\resizebox{.6\textwidth}{!}{%
%\resizebox{\textwidth}{!}{%
\renewcommand{\arraystretch}{1.1}
\label{ta4}
\begin{tabular}{c|cc|c|c|}
\cline{2-5}
                                   & \multicolumn{1}{c|} {Recall} & {Precision} & F1 score & Support \\ \hline
\multicolumn{1}{|c|}{0}            & \multicolumn{1}{c|} {0.98}   & {0.93}      & 0.95      & 52      \\ \hline
\multicolumn{1}{|c|}{1}            & \multicolumn{1}{c|} {0.94}   & {0.98}      & 0.96      & 67      \\ \hline
\multicolumn{1}{|c|}{Accuracy}     & \multicolumn{2}{c|}{}                   & 0.96      & 119     \\ \hline
\multicolumn{1}{|c|}{Macro avg.}    & \multicolumn{1}{c|} {0.96}   & {0.96}      & 0.96      & 119     \\ \hline
\multicolumn{1}{|c|}{Weighted avg.} & \multicolumn{1}{c|} {0.96}   & {0.96}      & 0.96      & 119     \\ \hline
\end{tabular}
}
\end{table}

\begin{figure*}[htbp]
%\label{fig: loss of best model}
\centerline{\includegraphics[width=.7\textwidth]{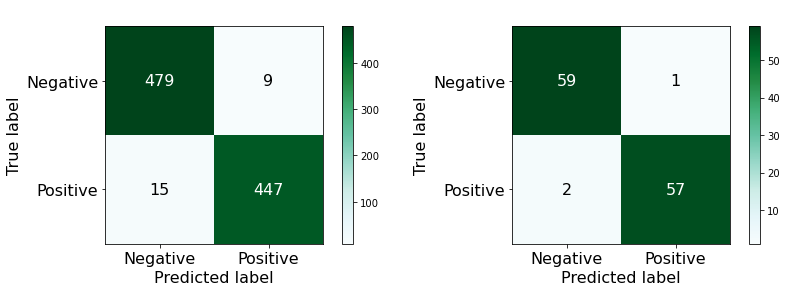}}
\caption{Training (left) and testing (right) confusion matrices for the transfer learning-based Inceptionv3 model.}
\label{inc1}
\end{figure*}

\begin{figure}[htbp]
%\label{fig: loss of best model}
\centerline{\includegraphics[width=0.5\textwidth]{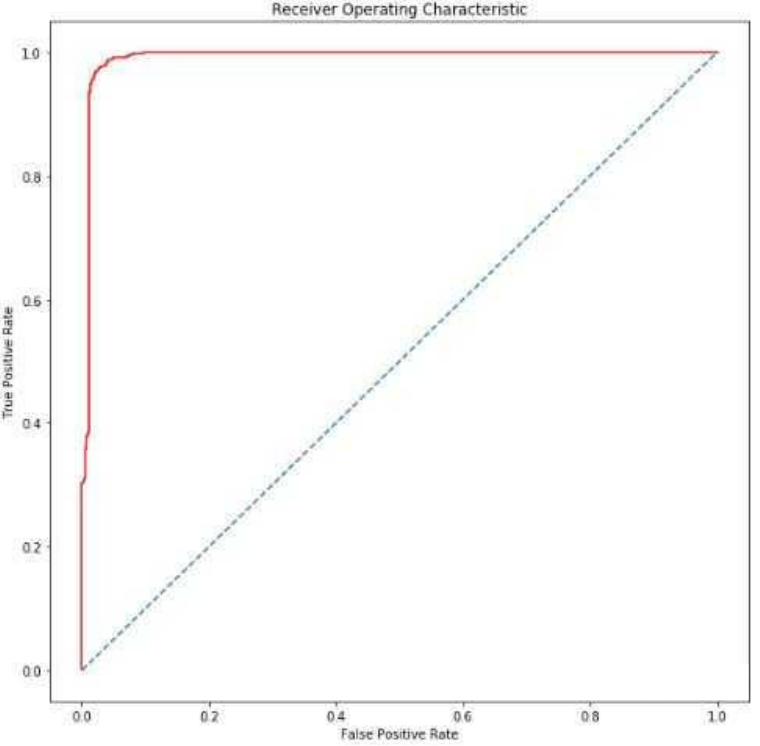}}
\caption{ROC curve for the transfer learning based on Inceptionv3.}
\label{r99}
\end{figure}

\begin{figure}[htbp]
%\label{fig: loss of best model}
\centerline{\includegraphics[width=.7\textwidth]{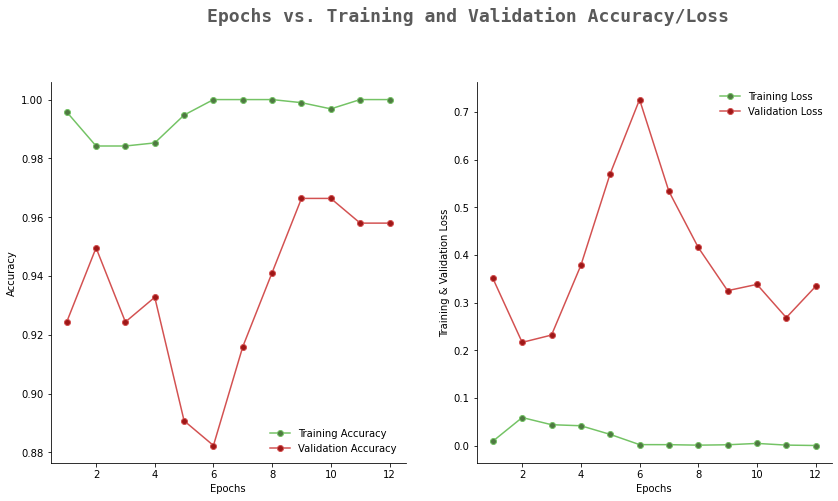}}
\caption{Training progress curve for the transfer learning-based Inceptionv3 model.}
\label{inc2}
\end{figure}

\begin{figure}[htbp]
%\label{fig: loss of best model}
\centerline{\includegraphics[width=.7\textwidth]{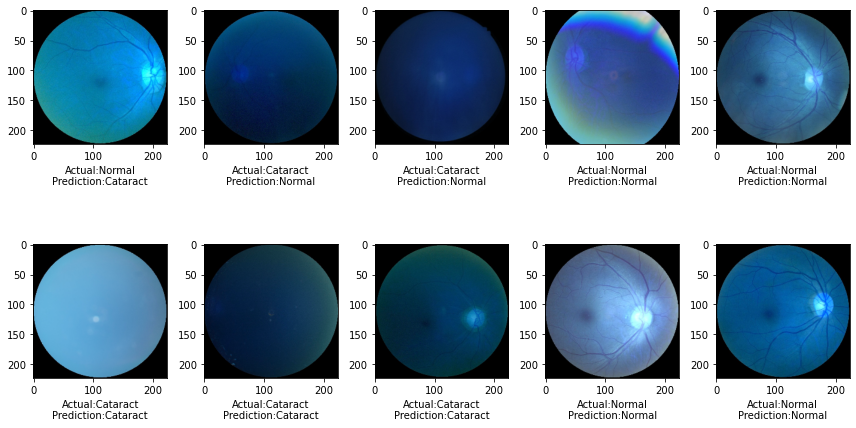}}
\caption{Sample results for the transfer learning-based Inceptionv3 model.}
\label{inc3}
\end{figure}

We also evaluated the effectiveness of the InceptionV3 transfer learning model in diagnosing diabetic retinopathy. This model was trained on a collection of retinal fundus images, and you can find the detailed classification report in Table \ref{tainc}. The confusion matrices for the validation set and test set are illustrated in Figure \ref{inc1}. The InceptionV3 model correctly classified 479 images as positive for diabetic retinopathy, while it incorrectly identified 447 images as positive for diabetic retinopathy when they were actually negative. Furthermore, there were 9 false negatives and only 15 true negatives. The model accurately classified 59 images as positive for diabetic retinopathy, while incorrectly classifying 1 image as positive for diabetic retinopathy when it was actually negative. Additionally, there were 2 false negatives and 57 true negatives. Figure \ref{r99} presents the ROC curve for the transfer learning of the Inceptionv3 model. Figure \ref{inc2} depicts the training and validation progress curve for accuracy and loss. Despite some misclassifications, the results indicate that the InceptionV3 transfer learning-based model is efficient in detecting diabetic retinopathy in retinal fundus images. However, some misclassified images by the model are shown in Figure \ref{inc3}.

\begin{table}[]
\centering
\caption{Classification report of transfer learning-based Inceptionv3 model.}
\resizebox{.7\textwidth}{!}{%
%\resizebox{\textwidth}{!}{%
\renewcommand{\arraystretch}{1.1}
\begin{tabular}{c|cc|c|c|}
\cline{2-5}
                                   & \multicolumn{1}{c|}{Recall} & Precision & F1 score & Support \\ \hline
\multicolumn{1}{|c|}{0}            & \multicolumn{1}{c|}{0.94}      & 1.00   & 0.97      & 65      \\ \hline
\multicolumn{1}{|c|}{1}            & \multicolumn{1}{c|}{1.00}      & 0.93   & 0.96      & 54      \\ \hline
\multicolumn{1}{|c|}{Accuracy}     & \multicolumn{2}{c|}{}                   & 0.97      & 119     \\ \hline
\multicolumn{1}{|c|}{Macro avg.}    & \multicolumn{1}{c|}{0.97}      & 0.97   & 0.97      & 119     \\ \hline
\multicolumn{1}{|c|}{Weighted avg.} & \multicolumn{1}{c|}{0.97}      & 0.97   & 0.97      & 119     \\ \hline
\end{tabular}
}
\label{tainc}
\end{table}


\subsection{Results for DiaCNN proposed Model}
\begin{figure*}[htbp]
%\label{fig: loss of best model}
\centerline{\includegraphics[width=.7\textwidth]{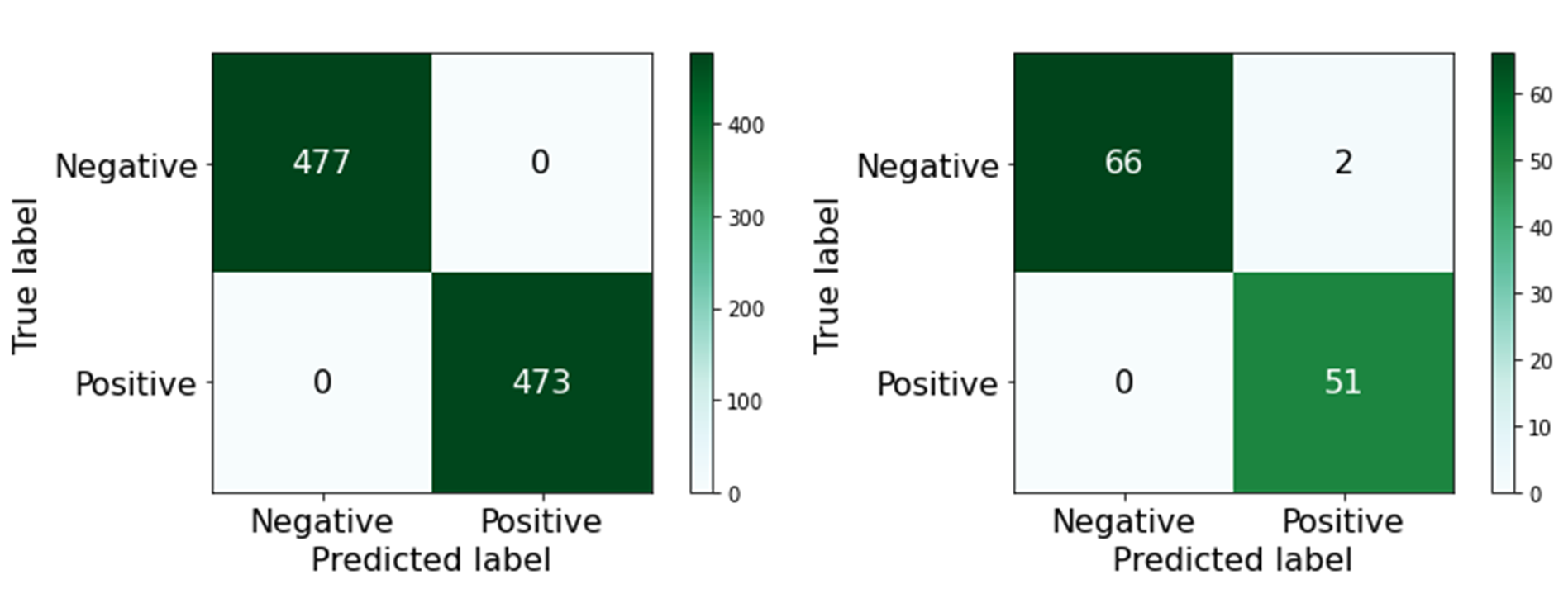}}
\caption{Training (left) and testing (right) confusion matrices for the DiaCNN proposed model with net-width 16.}
\label{f5}
\end{figure*}

\begin{figure}[htbp]
%\label{fig: loss of best model}
\centerline{\includegraphics[width=0.5\textwidth]{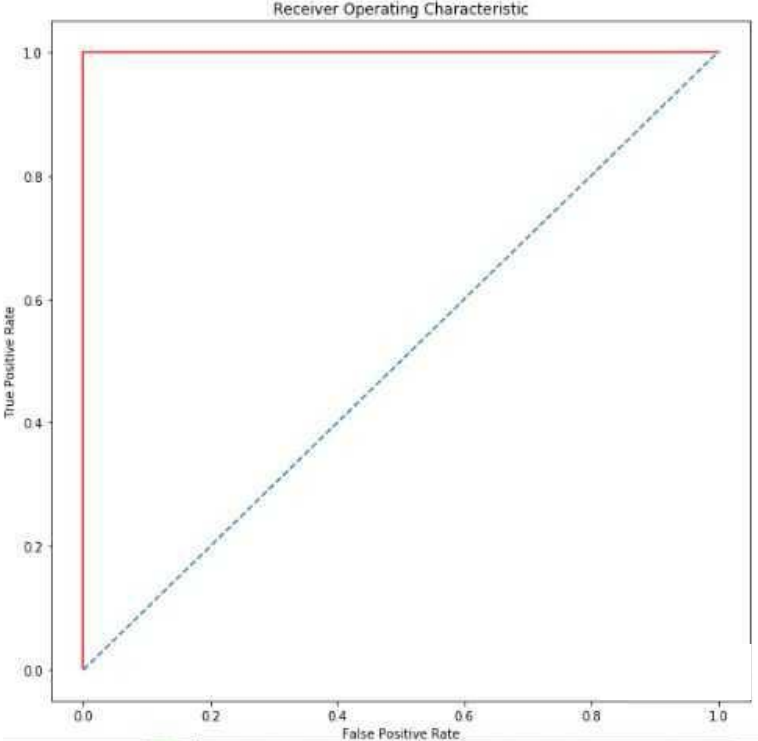}}
\caption{ROC curve for DiaCNN proposed model with net-width 16.}
\label{r100}
\end{figure}

The results of the proposed DiaCNN model with a net width of 16 on the ODIR dataset are presented in Table \ref{t4}. The classification report encompasses various performance metrics, including accuracy, sensitivity, specificity, precision, and F1 score. The model exhibited exceptional performance, achieving a 98\% score in all these metrics. The confusion matrices for the validation and testing phases are shown in Figure \ref{f5}. Figure \ref{r100} represents the ROC curve for the DiaCNN model with a net width of 16. The model demonstrated a strong ability to correctly identify positive cases of diabetic retinopathy, with high true positive rates and minimal false negatives, underscoring its effectiveness. The confusion matrices also show low false positive rates, indicating that the model did not incorrectly classify many negative cases as positive. Figure \ref{f7} illustrates a sample of output results from the model. The output results demonstrate the model's ability to accurately classify diabetic retinopathy cases in retinal fundus images. Overall, the DiaCNN model with net width 16 achieved high performance on the ODIR dataset and demonstrated its effectiveness in detecting diabetic retinopathy.

\begin{table}[h]
\centering
\caption{Classification report of DiaCNN proposed model with net-width 16.}
\resizebox{.7\textwidth}{!}{%
%\resizebox{\textwidth}{!}{%
\renewcommand{\arraystretch}{1.1}
\begin{tabular}{|c|c|c|c|c|}
\hline
 & Recall & Precision & F1 score  & Support \\
\hline
0 & 0.98 & 0.98 & 0.98 & 55 \\
\hline
1 & 0.98 & 0.98 & 0.98 & 64 \\
\hline
Accuracy &  &  & 0.98 & 119 \\
\hline
Macro Avg. & 0.98 & 0.98 & 0.98 & 119 \\
\hline
Weighted Avg. & 0.98 & 0.98 & 0.98 & 119 \\
\hline
\end{tabular}%
}
\label{t4}
\end{table}

%\begin{figure}[htbp]
%\label{fig: loss of best model}
%\centerline{\includegraphics[width=1\columnwidth]{figures/Picture7.png}}
%\caption{ROC curves for the DiaCNN proposed model with net-width 16.}
%\label{f6}
%\end{figure}
\begin{figure}[htbp]
%\label{fig: loss of best model}
\centerline{\includegraphics[width=.7\textwidth]{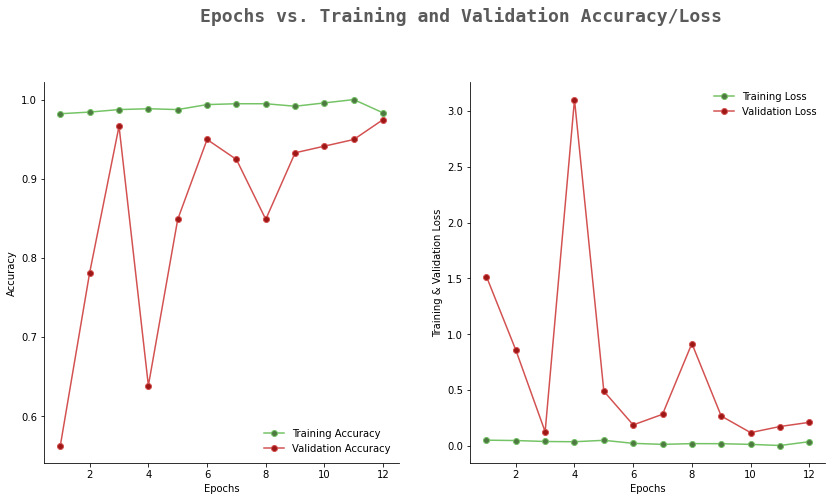}}
\caption{Training progress curve for the DiaCNN proposed model with net-width 16 models.}
\label{d16}
\end{figure}
\begin{figure}[htbp]
%\label{fig: loss of best model}
\centerline{\includegraphics[width=.7\textwidth]{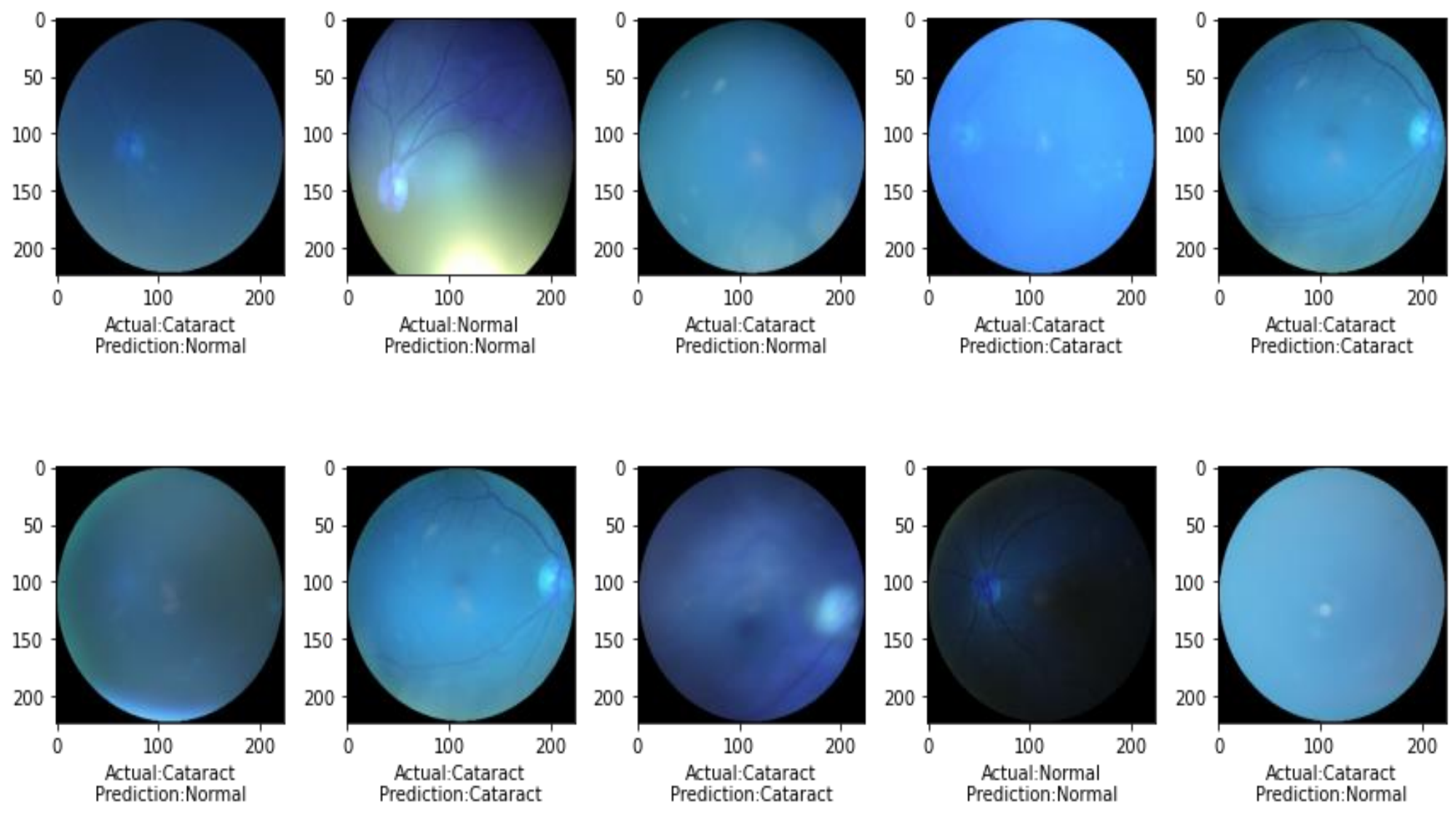}}
\caption{Sample results of the DiaCNN proposed model with net-width 16.}
\label{f7}
\end{figure}

\begin{figure*}[htbp]
%\label{fig: loss of best model}
\centerline{\includegraphics[width=.7\textwidth]{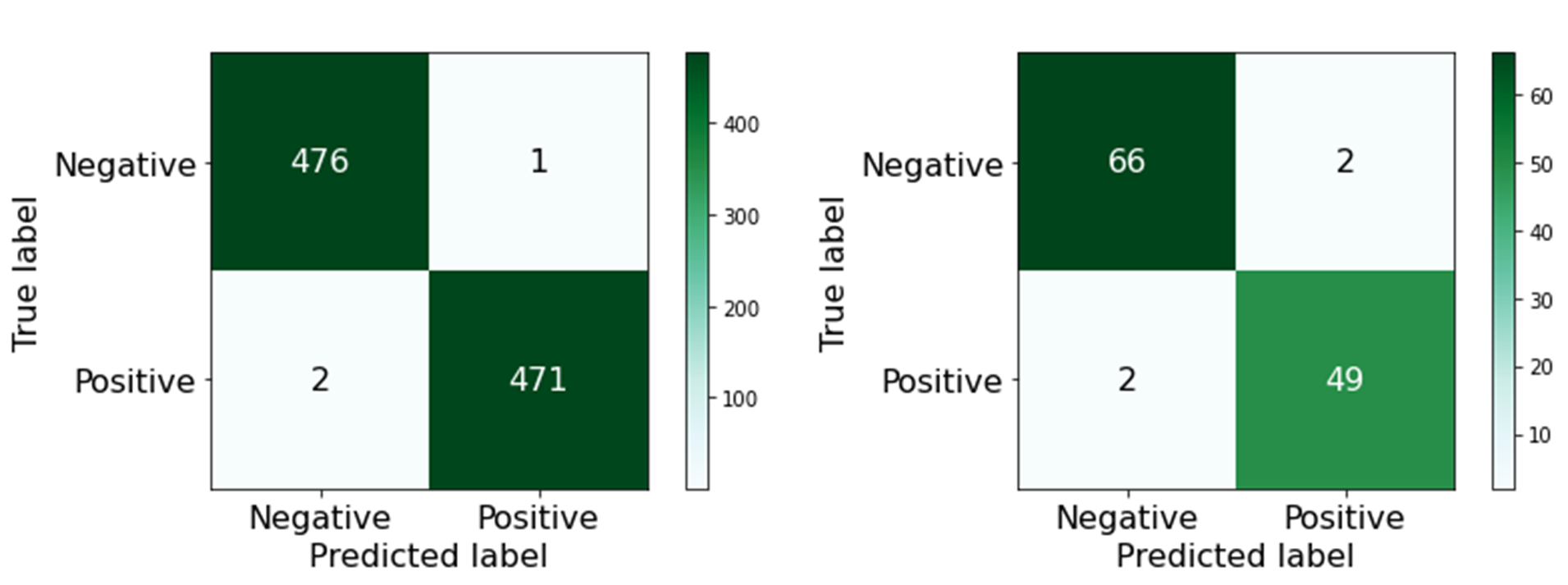}}
\caption{Training (left) and testing (right) Confusion matrices for the DiaCNN proposed model with net-width 12.}
\label{f8}
\end{figure*}

\begin{figure}[htbp]
%\label{fig: loss of best model}
\centerline{\includegraphics[width=0.5\textwidth]{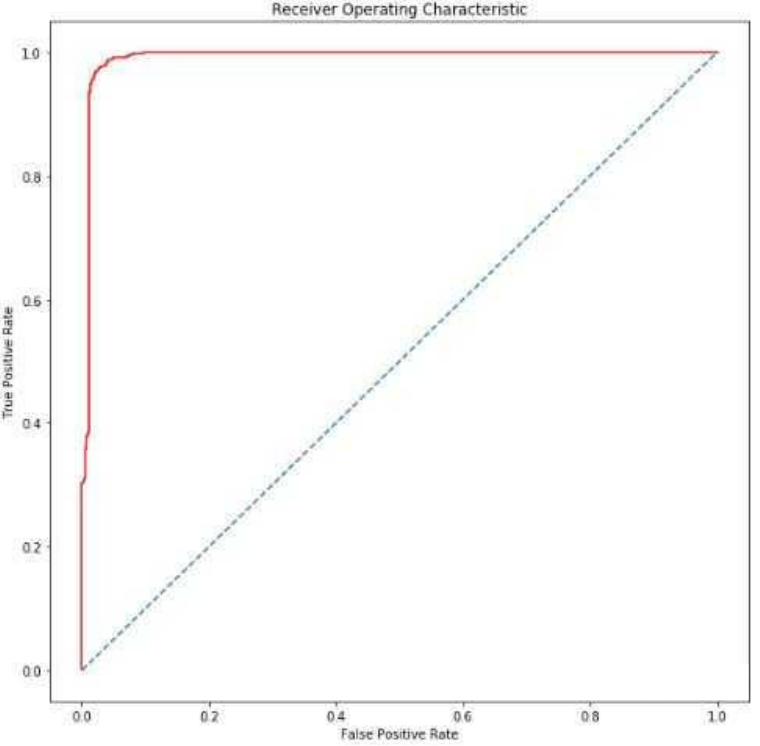}}
\caption{ROC curve for DiaCNN proposed model with net-width 12.}
\label{r9}
\end{figure}

\begin{figure}[htbp]
%\label{fig: loss of best model}
\centerline{\includegraphics[width=.7\textwidth]{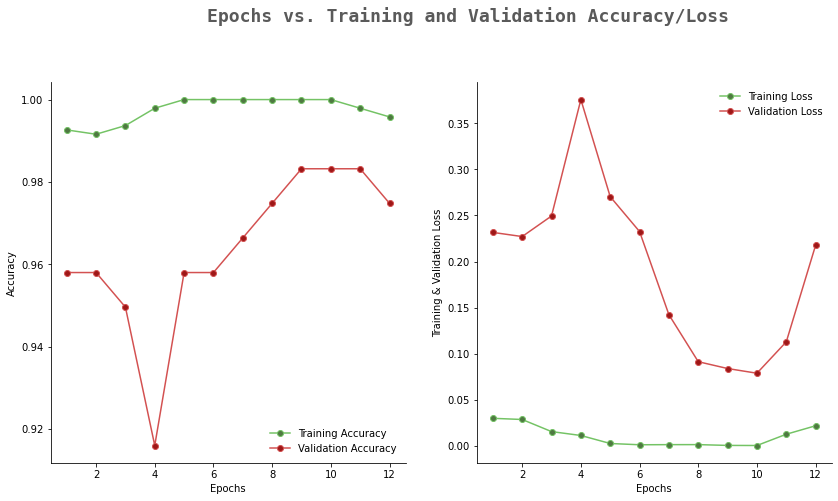}}
\caption{Training progress curve for the DiaCNN proposed model with net-width 12 models.}
\label{d12}
\end{figure}

\begin{figure}[htbp]
%\label{fig: loss of best model}
\centerline{\includegraphics[width=.7\textwidth]{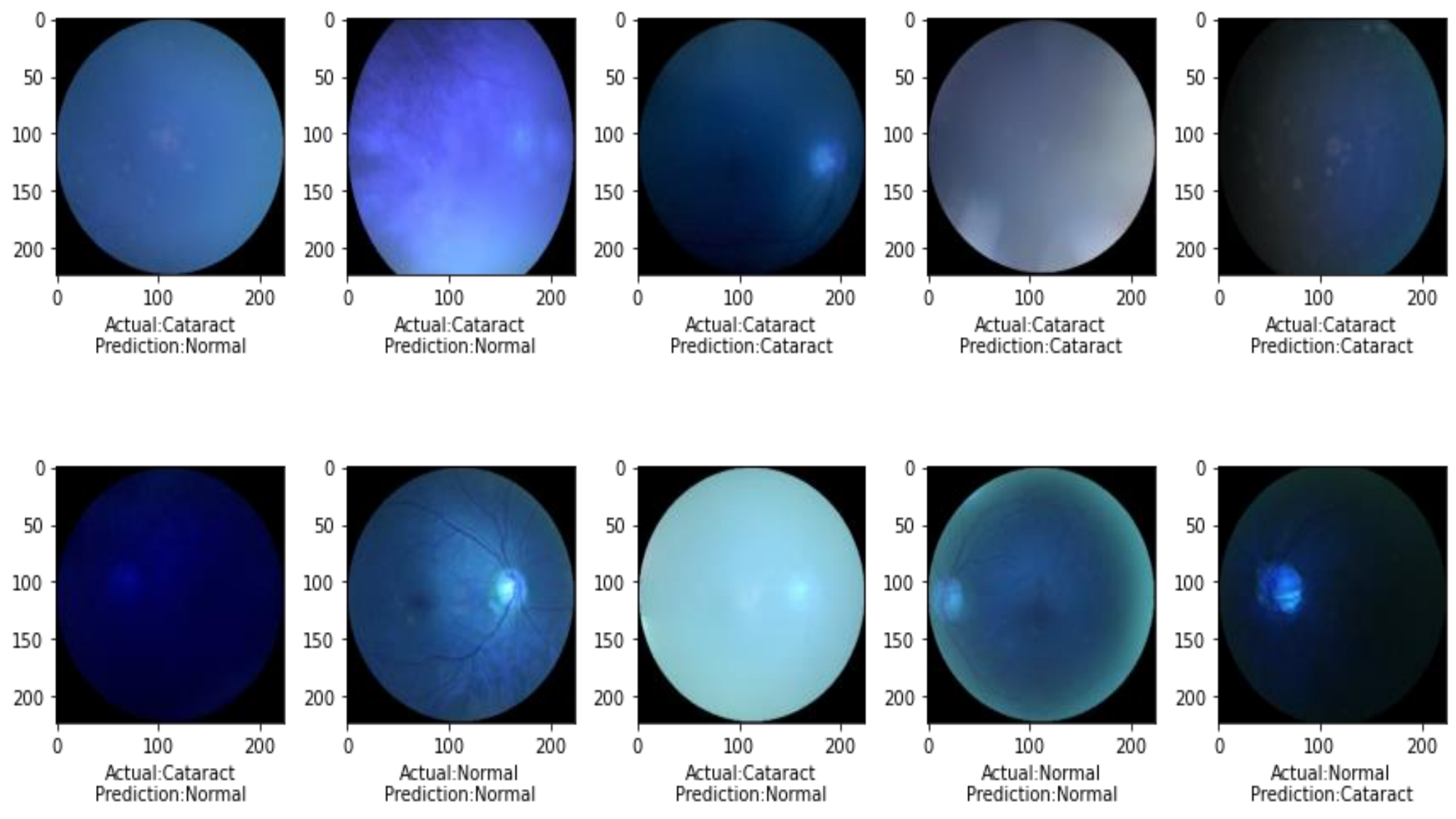}}
\caption{Sample results of the DiaCNN proposed model with net-width 12.}
\label{f10}
\end{figure}

Table \ref{t5} shows the classification report findings of the second proposed classification model for the ODIR dataset in terms of accuracy, sensitivity, specificity, precision, and F1 score using DiaCNN model with a net width of 12. This model provides a 97\%, 97\%, 97\%, 97\%, and 97\% for accuracy, sensitivity, specificity, precision, and F1 score, respectively. Figure \ref{f8} also shows the confusion matrices for the validation and testing phases. Figure \ref{r9} presents the ROC curve for DiaCNN proposed model with net-width 12. Training progress curve for the DiaCNN proposed model with net-width 12 models is presented in Figure \ref{d12}. Figure \ref{f10} also depicts an example of output results.

\begin{table}[h]
\centering
\caption{Classification report of DiaCNN proposed model with net-width 12.}
\resizebox{.7\textwidth}{!}{%
%\resizebox{\textwidth}{!}{%
\renewcommand{\arraystretch}{1.1}
\begin{tabular}{|c|c|c|c|c|}
\hline
 & Recall & Precision & F1 score  & Support \\
\hline
0 & 0.96 & 0.98 & 0.97 & 55 \\
\hline
1 & 0.98 & 0.97 & 0.98 & 64 \\
\hline
Accuracy &  &  & 0.97 & 119 \\
\hline
Macro Avg. & 0.97 & 0.98 & 0.97 & 119 \\
\hline
Weighted Avg. & 0.97 & 0.97 & 0.97 & 119 \\
\hline
\end{tabular}%
}
\label{t5}
\end{table}

%\begin{figure}[htbp]
%\label{fig: loss of best model}
%\centerline{\includegraphics[width=1\columnwidth]{figures/Picture10.png}}
%\caption{ROC curves for the Transfer learning based on Inceptionv3.}
%\label{f9}
%\end{figure}

\section{Discussion and comparison with the state-of-the-art methods}
Table \ref{total} summarizes the evaluation results obtained for the proposed models. Four different models have been trained and tested, including transfer learning-based models using InceptionResNetv2 and Inceptionv3 architectures and two versions of the DiaCNN model with net width 12 and 16. The table shows the evaluation metrics for both the training and testing phases, including accuracy, sensitivity, specificity, precision, and F1 score. The results demonstrate that the Inceptionv3-based transfer learning model performed remarkably well, achieving outstanding metrics in terms of accuracy, sensitivity, specificity, precision, and F1 score during both the training and testing phases. In the training phase, it reached an accuracy of 99.7\%, sensitivity of 99.4\%, specificity of 100\%, precision of 99.7\%, and F1 score of 100\%. This superior performance is evident as Inceptionv3 outperformed all other models in these metrics for the training data. However, in the testing phase, the DiaCNN model with a net width of 16 emerged as the top performer, achieving an impressive accuracy of 98.3\%, 100\% sensitivity and specificity, precision of 96.2\%, and an F1 score of 98.1\%. This indicates that the DiaCNN model with a net width of 16 has the potential to generalize well on unseen data. The proposed models' success can be credited to advanced deep learning architectures and transfer learning techniques that enable them to learn distinguishing features and decrease overfitting. These models can provide an accurate and dependable diagnosis of diabetic retinopathy, indicating an improvement in healthcare quality for patients with this ailment.
% \usepackage{multirow}
\begin{table*}[]
\centering
\caption{The obtained results for the training and testing phases of the proposed models.}
\resizebox{.7\textwidth}{!}{%
%\resizebox{\textwidth}{!}{%
\renewcommand{\arraystretch}{1.1}
\begin{tabular}{|c|cccccccccc|}
\hline
\multirow{3}{*}{Models}                        & \multicolumn{10}{c|}{Evaluation   metrics}                                                                                                                                                                                                                                   \\ \cline{2-11} 
                                               & \multicolumn{5}{c|}{Training}                                                                                                                   & \multicolumn{5}{c|}{Testing}                                                                                               \\ \cline{2-11} 
                                               & \multicolumn{1}{c|}{Acc}  & \multicolumn{1}{c|}{Sen}  & \multicolumn{1}{c|}{Spec}  & \multicolumn{1}{c|}{Preci} & \multicolumn{1}{c|}{F1 score} & \multicolumn{1}{c|}{Acc}  & \multicolumn{1}{c|}{Sen}  & \multicolumn{1}{c|}{Spec}  & \multicolumn{1}{c|}{Preci} & F1 score \\ \hline
Transfer learning-based InceptionResNetv2 model & \multicolumn{1}{c|}{97.5} & \multicolumn{1}{c|}{98.0} & \multicolumn{1}{c|}{96.8} & \multicolumn{1}{c|}{97.4}  & \multicolumn{1}{c|}{97.0}      & \multicolumn{1}{c|}{97.5} & \multicolumn{1}{c|}{98.3} & \multicolumn{1}{c|}{96.6} & \multicolumn{1}{c|}{97.4}  & 96.7      \\ \hline
Transfer learning-based Inceptionv3 model        & \multicolumn{1}{c|}{99.7} & \multicolumn{1}{c|}{99.4} & \multicolumn{1}{c|}{100}  & \multicolumn{1}{c|}{99.7}  & \multicolumn{1}{c|}{100}       & \multicolumn{1}{c|}{97.5} & \multicolumn{1}{c|}{94.6} & \multicolumn{1}{c|}{100}  & \multicolumn{1}{c|}{97.2}  & 100       \\ \hline
DiaCNN model with net-width 12                 & \multicolumn{1}{c|}{99.7} & \multicolumn{1}{c|}{99.6} & \multicolumn{1}{c|}{99.6} & \multicolumn{1}{c|}{99.8}  & \multicolumn{1}{c|}{99.7}      & \multicolumn{1}{c|}{96.6} & \multicolumn{1}{c|}{96.1} & \multicolumn{1}{c|}{97.1} & \multicolumn{1}{c|}{96.1}  & 96.1      \\ \hline
DiaCNN model with net-width 16                 & \multicolumn{1}{c|}{100}  & \multicolumn{1}{c|}{100}  & \multicolumn{1}{c|}{100}  & \multicolumn{1}{c|}{100}   & \multicolumn{1}{c|}{100}       & \multicolumn{1}{c|}{98.3} & \multicolumn{1}{c|}{100}  & \multicolumn{1}{c|}{100}  & \multicolumn{1}{c|}{96.2}  & 98.1      \\ \hline
\end{tabular}
}
\label{total}
\end{table*}
This work provides two different frameworks for eye disease diagnosis based on the benefits of deep CNN models, especially those based on the transfer learning method. Firstly, Cataract and Normal information are extracted from the Dataset; the number of images in the left cataract is 304 images, and the number of images in the right cataract is 290 images. After that, the datasets are Divided into features and targets. Then, creating the CNN Model based on the transfer learning method which is based on two different pre-trained models, InceptionResNetv2 and Inceptionv3. These two frameworks have different architectures. The first one consists of 54,722,273 for trainable parameters, 60,544 non-trainable parameters, and a total number parameter of 54,782,817. Also, the second one consists of 22,345,505 trainable parameters, 34,432 non-trainable parameters, and a total number parameter of 22,379,937. Then the model is trained on the training dataset with a training parameter and random initialization weights, a batch size of 64, a learning rate of 0.00001, and several epochs of 20.
The dataset was split into two groups: 80\% for training and 20\% for testing. To prevent overfitting, the training data was further divided into training and validation sets.

\begin{figure}[htbp]
%\label{fig: loss of best model}
\centerline{\includegraphics[width=0.5\textwidth]{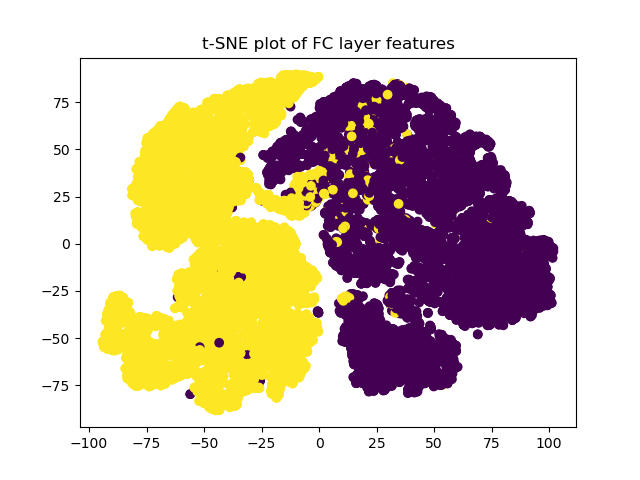}}
\caption{t-SNE graph for DiaCNN proposed model.}
\label{tn1}
\end{figure}

\begin{figure}[htbp]
%\label{fig: loss of best model}
\centerline{\includegraphics[width=0.5\textwidth]{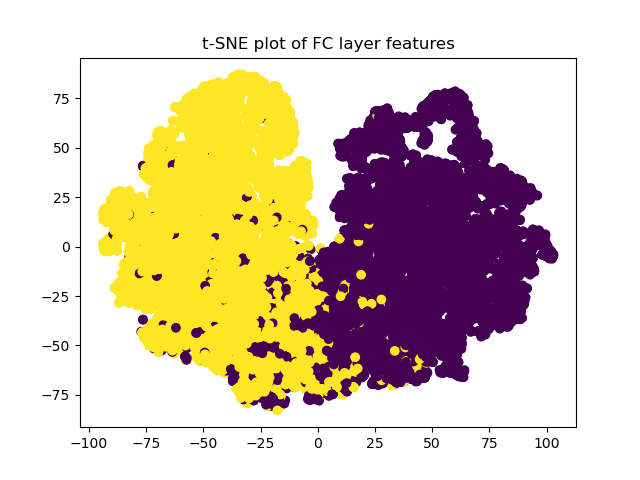}}
\caption{t-SNE graph for Inceptionv3 proposed model.}
\label{tn2}
\end{figure}

\begin{figure}[htbp]
%\label{fig: loss of best model}
\centerline{\includegraphics[width=0.5\textwidth]{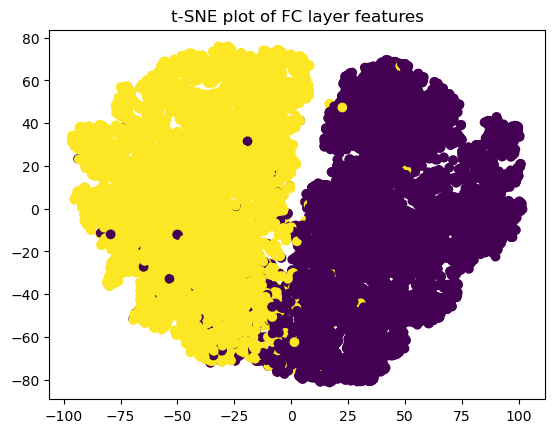}}
\caption{t-SNE graph for InceptionResNetv2 proposed model.}
\label{tn3}
\end{figure}

The t-SNE graph of the fully connected layers is a visualization technique that allows to explore and understand the distribution of data in high-dimensional space. The FC layers are usually positioned as the final network layers before the output layer, tasked with converting high-level features from earlier layers into predictions or classifications.
t-SNE, short for t-Distributed Stochastic Neighbor Embedding, is a technique for reducing the dimensionality of complex high-dimensional data, aiming to retain local data structure and relationships in a lower-dimensional space.
In the case of the FC layers, the t-SNE graph provides insights into how the extracted features are distributed and grouped within the network. It helps to visualize and analyze the patterns and similarities between the feature representations of different data samples.
By plotting the t-SNE graph of the FC layers, clusters or groupings of data points that have similar feature representations can be observed. Points that are close to each other in the t-SNE graph share similar characteristics or attributes. This visualization allows to identify potential patterns or separations in the data that may be useful for further analysis or decision-making.
Figures \ref{tn1}, \ref{tn2} and \ref{tn3} present the t-SNE graph for DiaCNN, Inceptionv3 and InceptionResNetv2 models. 


Finally, the proposed models are compared from the computational time of the testing phase as introduced in Table \ref{time}. By analyzing these results, we can observe the relative differences in testing time between the different models.
Comparing the transfer learning-based InceptionResNetv2 model and the Inceptionv3 model, we see that the Inceptionv3 model exhibits a shorter testing time of 198.7 seconds, which is approximately 20\% faster than the InceptionResNetv2 model testing time of 250.21 seconds. This improvement can be attributed to differences in the architectures and complexities of the two models. Inceptionv3 may have a more streamlined structure or more efficient operations, leading to faster computations during the testing phase.
Furthermore, comparing the DiaCNN model with net-width 12 and the transfer learning-based Inceptionv3 model, we observe that the DiaCNN model performs even better in terms of testing time, with a time of 101.54 seconds. This indicates a significant improvement of approximately 49\% compared to the Inceptionv3 model. The DiaCNN model's faster testing time can be attributed to its specific architecture, which may involve optimized operations or fewer computational layers.
Similarly, the DiaCNN model with a net-width 16 demonstrates a testing time of 154.78 seconds, which is an improvement compared to the InceptionResNetv2 model testing time, but slightly slower than the DiaCNN model with a net-width 12. This suggests that the DiaCNN model with a net-width 12 achieves a more favorable balance between computational efficiency and accuracy compared to the net-width 16 variants.
These observed improvements in testing time can be crucial in practical applications where efficiency is a priority. By reducing the time required for testing, the models can provide faster results, enabling timely diagnosis and decision-making. It is important to consider that the testing time can vary based on factors such as hardware specifications, optimization techniques, and the size and complexity of the models.

\begin{table}[]
\caption{Computational time of the examined approaches.}
\label{time}
\begin{tabular}{|c|c|}
\hline
\textbf{Model}                                           & \textbf{Testing time (s)} \\ \hline
\textbf{Transfer learning-based InceptionResNetv2 model} & \textbf{250.21}           \\ \hline
\textbf{Transfer learning-based Inceptionv3 model}       & \textbf{198.7}            \\ \hline
\textbf{DiaCNN model with net-width 12}                  & \textbf{101.54}           \\ \hline
\textbf{DiaCNN model with net-width 16}                  & \textbf{154.78}           \\ \hline
\end{tabular}
\end{table}

So, to put our results into context and demonstrate the efficacy of our proposed method, we compared our models with different existing state-of-the-art DR diagnostic models. Table \ref{tab1} compares several approaches proposed in different studies for detecting and classifying DR using various methodologies, algorithms, and performance metrics. Each approach is evaluated based on its accuracy, sensitivity, and specificity, among other metrics.
Overall, the proposed approaches seem to outperform the state-of-the-art, as they achieve high accuracy, sensitivity, and specificity in detecting and classifying DR. For instance, Neelu and Bhattacharya's approach achieved an accuracy of 96\%, a sensitivity of 90\%, and specificity of 94\%, which is impressive compared to the state-of-the-art. Similarly, Li and Yeh's approach achieved an accuracy of 91\%, sensitivity of 89\%, and specificity of 99\%, which is also very promising.
Among the proposed approaches, several methodologies and algorithms are used, including DNNs, siamese-like networks, DBNs, CNNs, cluster validity index and BAT optimization, KNN, SVM, and random forests (RF), among others.
The use of these methodologies and algorithms indicates that the proposed approaches are diverse, and the authors have taken different approaches to address the problem of DR detection and classification. This diversity of approaches is an advantage because it suggests that there is no one-size-fits-all approach to detecting and classifying DR, and researchers can use various methodologies and algorithms to achieve high accuracy, sensitivity, and specificity.
It is worth noting that some of the proposed approaches focused on detecting and classifying specific DR classes, such as mild, moderate, severe, and proliferative DR (P-DR). This approach allows for a more targeted and accurate diagnosis of DR, which is crucial for providing appropriate treatment to patients.
The comparison table provides valuable insights into the different approaches proposed in various studies for detecting and classifying DR using diverse methodologies and algorithms. The results indicate that the proposed approaches outperform the state-of-the-art and achieve high accuracy, sensitivity, and specificity, which is promising for the early and accurate detection of DR.

% \usepackage{multirow}
\begin{table*}[]
\centering
\caption{Comparison of the proposed approach with state-of-the-art methods}
\resizebox{\textwidth}{!}{%
\renewcommand{\arraystretch}{1.2}
		\label{tab1}
\begin{tabular}{|c|c|c|c|c|}
\hline
Authors                                                                         & Task                                                                                                      & Methodology                                                                                                                                                      & Algorithm                                                                                                                  & Performance   Metrics                                                                                                                                                                                \\ \hline
\begin{tabular}[c]{@{}c@{}}Neelu and\\  Bhattacharya  2020 \cite{b2}\end{tabular} & \begin{tabular}[c]{@{}c@{}}No DR, NP-DR,  \\  and P-DR classes\end{tabular}                               & DNN                                                                                                                                                              & \begin{tabular}[c]{@{}c@{}}PCA Firefly algorithm   \\ Adam Optimizer\end{tabular}                                          & \begin{tabular}[c]{@{}c@{}}Accuracy 96\%,   \\ Sensitivity 90\%, and \\ Specificity 94\%\end{tabular}                                                                                               \\ \hline
Yuan Luo,   Wenbin Ye. 2020 \cite{b3}                                              & \begin{tabular}[c]{@{}c@{}}No DR, Mild,  \\  Moderate, Severe,\\  and P-DR classes\end{tabular}           & Siamese-like                                                                                                                                                     & Inceptionv3                                                                                                               & \begin{tabular}[c]{@{}c@{}}Accuracy 94\%,  \\ Sensitivity 82\%, \\ and Specificity 70\%\end{tabular}                                                                                                \\ \hline
Ambaji S.   Jadhav et al. 2020 \cite{b4}                                         & \begin{tabular}[c]{@{}c@{}}Normal DR,   Earlier DR, \\ Moderate DR,\\  and Severe DR classes\end{tabular} & DBN                                                                                                                                                              & MGS-ROA                                                                                                                    & \begin{tabular}[c]{@{}c@{}}Accuracy 93.1\%, \\ Sensitivity 86.3\%, and \\ Specificity 95.4\%\end{tabular}                                                                                         \\ \hline
Shu-I Pao et   al. 2020 \cite{b5}                                                & \begin{tabular}[c]{@{}c@{}}No DR, Mild   DR, \\ and Severe DR classes\end{tabular}                        & CNN                                                                                                                                                              & Bichannel CNN   model                                                                                                      & \begin{tabular}[c]{@{}c@{}}Accuracy 87.8\%, \\ Sensitivity 77.8\%, \\ and Specificity 93.88\%\end{tabular}                                                                                        \\ \hline
Cheruku et al. 2017 \cite{b6}                                                     & \begin{tabular}[c]{@{}c@{}}No DR, Mild   DR, \\ and Severe DR classes\end{tabular}                        & \begin{tabular}[c]{@{}c@{}}Cluster   Validity Index and BAT\\  Optimization with \\ Novel Fitness Function\end{tabular}                                          & \begin{tabular}[c]{@{}c@{}}Conventional   RBFN, \\ RBFN + Ratio Index,\\  RBFN + DunnIndex,\\  RBFN + DVIndex\end{tabular} & \begin{tabular}[c]{@{}c@{}}Conventional   RBFN, RBFN\\  + Ratio Index, RBFN + \\ DunnIndex, RBFN + \\ DVIndex archived of   \\ Accuracy (\%) 68.53,\\  70.00, 69.33, 69.56, respectively\end{tabular} \\ \hline
Revathy R et   al, 2019 \cite{b20}                                             & DR, NDR                                                                                                   & KNN, SVM, RF                                                                                                                                                     & The hybrid   model of all 3                                                                                                & Accuracy 82\%                                                                                                                                                                                       \\ \hline
Zhu et al, 2019 \cite{b21}                                                       & DR, NDR                                                                                                   & NB SVM   classifier                                                                                                                                              & NB SVM   classifier                                                                                                        & \begin{tabular}[c]{@{}c@{}}Accuracy 80\%, \\ Sensitivity 100\%, \\ and Specificity 67\%\end{tabular}                                                                                              \\ \hline
Li and Yeh,2019 \cite{b22}                                                       & \begin{tabular}[c]{@{}c@{}}NP-DR (level   1,2) \\ P-DR (level 1,2)\end{tabular}                           & DCNN                                                                                                                                                             & \begin{tabular}[c]{@{}c@{}}Fractional   max pooling \\ SVM with TLBO\end{tabular}                                          & \begin{tabular}[c]{@{}c@{}}Accuracy 91\%, \\ Sensitivity 89\%, and\\  Specificity 99\%\end{tabular}                                                                                               \\ \hline
LI and GUO, 2019 \cite{b23}                                                       & \begin{tabular}[c]{@{}c@{}}No DR NPDR   \\ Mild PDR Severe PDR\end{tabular}                               & DCNN                                                                                                                                                             & \begin{tabular}[c]{@{}c@{}}Inception v3\\    \\ ResNet\end{tabular}                                                        & Accuracy 89\%                                                                                                                                                                                       \\ \hline
Joel J, J   Kivinen, 2019 \cite{b24}                                             & \begin{tabular}[c]{@{}c@{}}Referable and   nonreferable\\  DR (R-DR, NRDR)\end{tabular}                   & DCNN                                                                                                                                                             & Inceptionv3                                                                                                               & \begin{tabular}[c]{@{}c@{}}Accuracy 91\%, \\ Sensitivity 85\%, \\ and Specificity 96\%\end{tabular}                                                                                               \\ \hline
Jadoon   et al, 2019 \cite{b25}                                                   & \begin{tabular}[c]{@{}c@{}}NPDR Mild   Moderate \\ Severe PDR\end{tabular}                                & DCNN                                                                                                                                                             & \begin{tabular}[c]{@{}c@{}}Resnet50   Inceptionv3, \\ Dense-121 Dense169\end{tabular}                                    & \begin{tabular}[c]{@{}c@{}}Accuracy 80.8\%\\ and Specificity 86.7\%\end{tabular}                                                                                                                 \\ \hline
Nikhil  and Angel 2019 \cite{b26}                                                & \begin{tabular}[c]{@{}c@{}}NPDR Mild   Moderate \\ Severe PDR\end{tabular}                                & DCNN                                                                                                                                                             & AlexNet,   Vgg16, InceptionV3                                                                                          & Accuracy 80.1\%                                                                                                                                                                                   \\ \hline
Dutta   et al, 2018 \cite{b27}                                                   & \begin{tabular}[c]{@{}c@{}}Mild NPR   Moderate \\ NPR Severe PR\end{tabular}                              & BNN, DNN, CNN   (VGGNet)                                                                                                                                         & VGGNet (CNN)                                                                                                               & Accuracy 78.3\%                                                                                                                                                                                   \\ \hline
Feng Li et   al, 2019 \cite{b28}                                                  & \begin{tabular}[c]{@{}c@{}}No DR Mild DR   NPDR \\ Severe NPDR PDR\end{tabular}                           & Deep transfer   learning                                                                                                                                         & Inceptionv3                                                                                                               & \begin{tabular}[c]{@{}c@{}}Accuracy 93.49\%, \\ Sensitivity 96.93\%,\\  and Specificity 93.45\%\end{tabular}                                                                                      \\ \hline
Li et al. (2023) \cite{n1}                                                       & R-DR, NRDR                                                                                                & Multimodal Information Fusion                                                                                                                                    & \begin{tabular}[c]{@{}c@{}}different fusion methods with\\  different Deep Learning Backbone\end{tabular}                  & \begin{tabular}[c]{@{}c@{}}Accuracy 91.1\%\\ Sensitivity 86\%, \\ and Specificity 88\%\end{tabular}                                                                                                            \\ \hline
Zang et al. (2022) \cite{n2}                                                     & DR, NDR                                                                                                   & Biomarker Activation Map (BAM)                                                                                                                                   & Generative adversarial learning                                                                                            & \begin{tabular}[c]{@{}c@{}}F1 score 0.63 ± 0.08,\\    \\ Precision 0.64 ± 0.16,\\    \\and Recall 0.65 ± 0.08\end{tabular}                                                                                \\ \hline
Basu, Soham et al. (2022) \cite{n3}                                             & R-DR, NRDR                                                                                                & \begin{tabular}[c]{@{}c@{}}Segmentation of Blood Vessels, \\ Optic Disc Localization, Detection of   Exudates \\ and Diabetic Retinopathy Diagnosis\end{tabular} & \begin{tabular}[c]{@{}c@{}}DCNN, k-means clustering, \\ contour   detection\end{tabular}                                   & \begin{tabular}[c]{@{}c@{}}Vessel segmentation: \\ 95.93\% Accuracy; \\ Optic disc localization:\\  98.77\%   Accuracy; \\ DCNN: 75.73\% Accuracy\end{tabular}                                       \\ \hline
\multirow{3}{*}{Proposed Approach}                                              & \multirow{3}{*}{Normal  and abnormal}                                                                     & \multirow{2}{*}{Deep transfer   learning}                                                                                                                        & InceptionResNetv2                                                                                                          & Accuracy 98.32\%                                                                                                                                                                                     \\ \cline{4-5} 
                                                                                &                                                                                                           &                                                                                                                                                                  & Inceptionv3                                                                                                                & Accuracy 97.48\%                                                                                                                                                                                     \\ \cline{3-5} 
                                                                                &                                                                                                           & Residual learning-based model                                                                                                                                    & DiaCNN                                                                                                                     & Accuracy 100\%                                                                                                                                                                                       \\ \hline
\end{tabular}
}
\end{table*}

\section{Conclusion and Future Works }
This research presents an innovative methodology employing deep learning strategies to achieve accurate diagnoses of diabetic retinopathy and other ocular ailments. Performance evaluation conducted on the Ocular Disease Intelligent Recognition dataset yielded outstanding outcomes during the training, testing, and validation stages. Specifically, the InceptionResNetv2 model, harnessing the power of transfer learning, recorded accuracies of 97.5\% in both training and testing. Simultaneously, the Inceptionv3 model achieved an exemplary 99.7\% training accuracy and 97.5\% testing accuracy. Notably, our custom-developed DiaCNN model demonstrated unparalleled precision with a perfect 100\% training accuracy and a commendable 98.3\% during testing. These findings emphasize the transformative potential of our approach to refine diagnostic precision for diabetic retinopathy and other eye conditions, paving the way for timely interventions and enhanced patient care. Nevertheless, it is crucial to recognize the study's constraints, with the size of the dataset being a potential influencer on diagnostic accuracy. Upcoming research endeavors should consider integrating contemporary methodologies such as data augmentation to amplify diagnostic precision. There exists a multitude of promising pathways for further enhancements in this domain. Crafting systems resilient to noise, harnessing feature fusion methodologies for superior classification, and delving into deep learning paradigms coupled with segmentation tactics stand out as promising future research directions. Such endeavors could rectify the current study's limitations and further bolster the reliability and accuracy of our proposed methodology. In essence, this research forms a foundational stepping stone towards the development of robust and efficient diagnostic platforms for diabetic retinopathy and a spectrum of eye diseases. By addressing pinpointed limitations and embarking on suggested advancements, we stand on the precipice of a transformative era in ocular disease diagnosis and management, offering a brighter vision for the future.
Thus, when contrasted with current state-of-the-art diagnostic methods, our models manifest a significant enhancement in classification accuracy. This research not only attests to the viability of incorporating advanced deep learning techniques, particularly transfer learning, into the realm of DR diagnosis, but also sheds light on the potential of such techniques in transforming the diagnostic landscape for other eye diseases.
So, the substantial improvements in accuracy offered by our models can herald expedited and more accurate interventions, which, in turn, can play a pivotal role in reducing the rate of DR-induced blindness. In conclusion, our research, underscored by its groundbreaking approach and outstanding results, presents a paradigm shift in DR diagnostic techniques, emphasizing the profound potential of deep learning in bolstering early detection and enhancing patient care.

\section{Acknowledgment}
The authors would like to acknowledge the Princess Nourah bint Abdulrahman University Researchers Supporting Project number (PNURSP2023R66), Princess Nourah bint Abdulrahman University, Riyadh, Saudi Arabia.

\section{Funding}
The authors would like to acknowledge the Princess Nourah bint Abdulrahman University Researchers Supporting Project number (PNURSP2023R66), Princess Nourah bint Abdulrahman University, Riyadh, Saudi Arabia.

 %\bibliographystyle{elsarticle-num} 
 %\bibliography{cas-refs}





% \end{thebibliography}
\end{document}